\documentclass{elsart}
\usepackage{latexsym,amsmath,amssymb,amsfonts,graphicx}
\numberwithin{equation}{section}
\numberwithin{figure}{section}

\newcommand{\nc}{\newcommand} 
\nc{\diff}[2]{\frac{d #1}{d #2}}
\nc{\diffn}[3]{\frac{d^{#3} #1}{d {#2}^{#3}}} 
\nc{\pdiff}[2]{\frac{\partial #1}{\partial #2}} 
\nc{\abs}[1] {\lvert #1 \rvert} 
\nc{\norm}[1] {\rVert #1 \rVert} 
\renewcommand{\d}{\delta}
\nc{\g}{\gamma} 
\nc{\D}{\partial} 
\nc{\nn}{\nonumber} 
\DeclareMathOperator{\sech}{sech}

\nc{\intinf}{\int_{-\infty}^{\infty}}
\nc{\cL}{{\mathcal L}}
\nc{\cA}{{\mathcal A}}
\nc{\cB}{{\mathcal B}}
\nc{\cF}{{\mathcal F}}
\nc{\cM}{{\mathcal M}}
\nc{\cN}{{\mathcal N}}
\nc{\cH}{{\mathcal H}}
\nc{\cI}{{\mathcal I}}
\nc{\cP}{{\mathcal P}}
\nc{\cS}{{\mathcal S}}
\nc{\defeq}{\stackrel{\rm def}{=}}
\def\beq#1\eeq{\begin{equation}#1\end{equation}}
\nc{\RR}{\mathbb R}

\begin{document}
\begin{frontmatter}
\title{Strong NLS Soliton-Defect Interactions\thanksref{KK}}
\thanks[KK]{Dedicated to Klaus Kirchg\"{a}ssner on the occasion of his 70th 
birthday: a belated gift.  } 

\author{Roy H. Goodman\corauthref{cor}}
\address{Mathematical Sciences Department, New Jersey Institute of Technology,
Newark, NJ 07102}
\ead{goodman@njit.edu}
\corauth[cor]{Corresponding author}

\author{Philip J. Holmes}
\address{Program in Applied and Computational Mathematics and 
Department of Mechanical and Aerospace Engineering,
Princeton University, Princeton, NJ 08544}
\ead{pholmes@princeton.edu}

\author{Michael I. Weinstein}
\address{Mathematical Sciences Research, Bell Laboratories--Lucent
Technologies, Murray Hill, NJ 07974}
\ead{miw@research.bell-labs.com}

\begin{abstract}

We consider the interaction of a nonlinear Schr\"odinger soliton with
a spatially localized (point) defect in the medium through which it
travels. Using numerical simulations, we find parameter regimes under
which the soliton may be reflected, transmitted, or captured by the
defect. We propose a mechanism of resonant energy transfer to a
nonlinear standing wave mode supported by the defect. Extending
Forinash et.\ al.~\cite{FPM:94}, we then derive a finite-dimensional model
for the interaction of the soliton with the defect via a collective
coordinates method. The resulting system is a three degree-of-freedom
Hamiltonian with an additional conserved quantity.  We study this
system both numerically and using the tools of dynamical systems
theory, and find that it exhibits a variety of interesting behaviors,
largely determined by the structures of stable and unstable manifolds
of special classes of periodic orbits. We use this geometrical
understanding to interpret the simulations of the finite-dimensional
model, compare them with the nonlinear Schr\"odinger simulations, and
comment on differences due to the finite-dimensional ansatz.

To fit into Arxiv's file size requirements, low-resolution
versions of certain large figures were used.  A version of this paper with the
full figures is available
at \begin{verbatim}http://m.njit.edu/~goodman/\end{verbatim}
\end{abstract}


\end{frontmatter}

\section{Introduction}
\label{s1}

In a previous study, involving the first and last authors~\cite{GSW},
(Bragg) resonant nonlinear propagation of light through optical
waveguides with a periodically varying refractive index profile and
localized defects was investigated. In that work, an approach to the
design of spatial defects in a periodic structure for the purpose of
trapping and localizing light pulses was suggested and explored.  The
technique involves resonant transfer of energy from traveling waves
(gap solitons) to {\it nonlinear} standing wave modes localized at the
defect. The cubic nonlinear Schr\"odinger (NLS) equation with
localized potentials, which we study in the present paper, provides a
simpler model exhibiting similar phenomena that is more amenable to
analysis  In particular, in numerical simulations of NLS solitons
incident on a single delta-well (point) defect in a one-dimensional
continuum, we find a variety of behaviors depending on the parameters
describing the soliton.

Several studies have examined the propagation of nonlinear waves through
variable or random media. The approach taken in~\cite{KGSV:90,Br:98} is to
view such a medium as sequence of individual {\it weak} scatterers, each
modeled by a repulsive delta function potential barrier. The interaction of a
soliton with an individual scatterer is formulated as a mapping of internal
soliton parameters; a soliton which interacts weakly with a scatterer adjusts
its internal parameters slightly due to radiative loss of energy.  The
interaction with the full medium is approximated by repeated composition of
this simple mapping.

The problem we address is different in a number of respects. We
consider {\it strong} interactions of a soliton incident on a {\it
single} scatterer or defect potential. Furthermore, our potential is
taken to be an {\it attractive} delta function potential well, which
has a single localized eigenstate (defect mode).  Therefore, an
incident soliton can be expected to break up into a soliton with
adjusted parameters, due to energy transfer to the localized defect
mode, and to outgoing radiation. Components of the soliton's energy
may be reflected by the potential, transmitted through the potential,
or captured by  its intrinsic modes.  These strong nonlinear
scattering interactions exhibit a great deal of complexity, which we
explore first by direct numerical simulation and then via finite
dimensional models.

In particular, we first  conduct a series of numerical experiments on
the partial differential equation (PDE), in which a variety of of
phenomena are observed, which we may partly explain by a resonant
transfer of energy to standing wave modes localized at the
defect. Second, we derive a finite-dimensional system of ordinary
differential equations (ODEs) that models the interaction of solitons
with nonlinear standing wave `defect' modes supported by the
potential. This part of the analysis is similar in spirit to our
earlier study of a finite dimensional reduction of the simpler case of
kinks interacting with a trapped mode in the sine-Gordon equation with
a point defect~\cite{GHW:01}. After reviewing the basic PDE model in
Section~{sec:pde} and \ref{sec:overview}, and describing the
results of direct numerical simulations in Sections~\ref{sec:dns}, we
outline the (formal) finite-dimensional reduction procedure in
Section~\ref{sec:ode}. We then describe in Section~\ref{sec:odesim} a
representative set of three numerical experiments, in the same
parameter ranges as the PDE studies, that reveal the kinds of soliton
transmission, reflection and transient capture behaviors that the ODEs
exhibit. Section~\ref{sec:ode_analysis} is devoted to analysis of the
ODEs. We describe invariant subspaces and special sets of orbits,
focusing on the stable and unstable manifolds of certain periodic
orbits. These are shown to partially `organize' the global dynamics,
in particular providing separatrices between transmitted and reflected
soliton orbits. In the final Section~\ref{sec:summary}, we make
comparisons between the PDE and ODE dynamics and
summarize. Detailed analytical calculations, and some background 
material, are relegated to an Appendix.

In addition to our specific study of NLS soliton-defect mode
interactions, we believe that the detailed comparison of PDE and ODE
solutions in this paper has general implications for similar
`collective coordinate' finite-dimensional representations commonly
used to study dynamical interactions of continuous fields.

\section{The nonlinear Schr\"odinger (NLS) equation with a point defect}
\label{sec:pde}

We consider a nonlinear Schr\"odinger (NLS) equation with a spatially localized
`attractive' impurity (a defect potential well)  at the origin:
\beq
iu_t+\frac{1}{2} u_{xx} + \abs{u}^2u + \g \; \d(x) u = 0,\ \ \g>0.
\label{eq:NLS}
\eeq
In the absence of a defect ($\g = 0$), this system supports a two-parameter
family of solitons of the form
\beq
u_{\rm{Sol}}(x,t) = \eta \sech(\eta(x - v t))e^{i(v x - \omega t)},
\label{eq:soliton}
\eeq 
where $\omega = -\frac{1}{2}(\eta^2 - v^2)$ is the temporal frequency.
Solitons are nonlinear bound states which  play a fundamental role in
the unperturbed ($\g=0$) NLS equation, and  we are especially
interested in their behavior in the perturbed system.  For $\g>0$,
(\ref{eq:soliton}) does not solve~\eqref{eq:NLS}, but far from the
defect,  the soliton propagates essentially without distortion and at
constant  speed $v$ due to the exponentially small overlap of soliton
and potential.

Equation~\eqref{eq:NLS} also supports {\it exact} nonlinear bound 
state or {\it defect mode} solutions of the form 
\beq 
u_{\rm{Def}}(x,t) = a e^{ia^2t/2}\sech{(a\abs{x} +\tanh^{-1}\frac{\g}{a})} ,
\label{eq:defectmode}
\eeq 
for all $a \ge \g$. These solutions are constructed from a stationary
($v=0$) soliton on each side of the defect pasted together at $x=0$ to satisfy
the conditions of continuity at $x=0$ and the appropriate jump condition in
the first derivative at $x=0$, $u(0^+)-u(0^-)=-2\g u(0)$. For both bound state
families, $u_{\rm Sol}$ and $u_{\rm Def}$, the frequency of oscillation
depends nonlinearly on its amplitude.

\begin{figure}
\begin{center}
\includegraphics[width=3in]{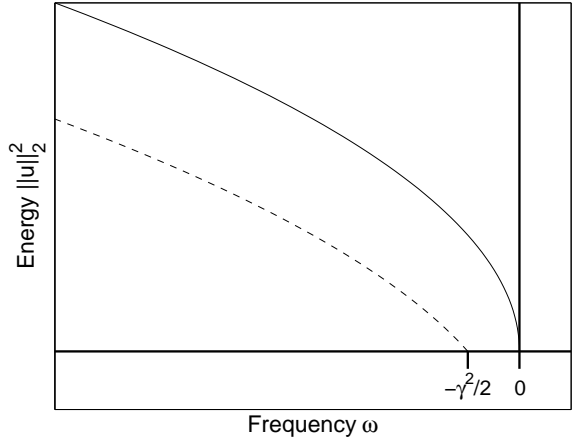}
\end{center}
\caption{The frequency and amplitude of a soliton in the absence
of a defect (solid), and of the defect mode (dashed).} \label{fig:L_v_freq}
\end{figure}

The chief concern of this paper is to understand strong interactions between
solitons and the delta well defect.  If the defect strength $\gamma$ is small
of the soliton velocity is large, then the interaction is weak: a small amount
of energy is lost to radiation, and the soliton continues past the defect with
minor changes to the parameters that define it. In the case of weak
interactions, estimates for the energy loss of the soliton can be obtained by
the first order perturbation theory (the Born
approximation)~\cite{Br:98,KGSV:90}.   When $\gamma$ is sufficiently large and
$v$ is small, stronger interactions can take place, the character of which may
be understood in terms of a nonlinear resonance that in some cases takes place
between the soliton and the defect mode.

Solitons with $v = 0$ have $\| u_{\rm{Sol}} \|_{L^2}^2 = 2\eta$ and frequency
$-\eta^2/2$, whereas nonlinear defect modes have $\| u_{\rm{Def}} \|_{L^2}^2 =
2(a-\g)$ and frequency $-a^2/2$.  In Figure~\ref{fig:L_v_freq} we plot the
squared $L^2$ norm of these two types of mode as functions of frequency. In
the following section we discuss the {\it bifurcation diagram} of
Figure~\ref{fig:L_v_freq} and its implications, and review the well-posedness
theory of the initial value problem and the dynamic stability theory for
$u_{\rm Sol}$ and $u_{\rm Def}$. Due to the relation between the square of
the $L^2$ norm with the electromagnetic energy in the context of optics, we
shall refer to $\| u_{\rm{Sol}} \|_{L^2}^2$ and $\| u_{\rm{Def}} \|_{L^2}^2$
as the {\it energy} of the soliton and defect modes, and more generally to
the square of the $L^2$ norm of $u$ over a region of space as the energy
contained in that region.

Note that no defect modes exist in
the frequency range $\omega \in (-\g^2/2, 0]$.  This observation is
crucial in predicting which solitons will be trapped and which
reflected by the potential.  We find, roughly, that sufficiently slow
solitons with $\eta > \g$ are trapped upon encountering the defect,
while slow solitons with $\eta < \g$ are reflected by the defect. This
suggests that trapping occurs via resonant energy transfer from the
soliton to the defect mode. If the incoming soliton has frequency less
than $-\g^2/2$, it can and may excite a nonlinear defect mode and
transfer its energy to that mode. Otherwise, as in~\cite{CM:95}, 
we find that the defect behaves as a scatterer, splitting the
incoming wave into transmitted, captured, and reflected parts.  This
behavior differs sharply from one seen in nonlinear optics: in the
nonlinear coupled mode equations describing the interaction of gap
solitons with nonlinear defect modes~\cite{GSW}, it was found that in
the absence of resonance, pulses were either coherently reflected or
transmitted after interacting with the defect, with little energy
captured or lost to radiation.

\section{Overview of well-posedness and stability}
\label{sec:overview}

\noindent {\it Structural properties of NLS:}
\ The nonlinear Schr\"odinger equation~\eqref{eq:NLS} is a Hamiltonian
system, which can be written in the form:
\beq
iu_t\ =\ \frac{\delta}{\delta u^*} \cH[u,u^*],\ \label{eq:NLS-hamform}
\eeq
where $\cH[u,u^*]$ denotes the Hamiltonian:
\begin{align}
\cH[u,u^*]\ &=\ 
\int\ \left(\ \frac{1}{2}|u_x|^2 \ -\ \frac{1}{2}|u|^4\ -\ \g\delta(x)|u|^2\ 
\right)  \ dx\nonumber\\
 &=\ 
 \int\ \left(\ \frac{1}{2}|u_x|^2 \ -\ \frac{1}{2}|u|^4\ \right)\ dx -\ 
 \g |u(0)|^2\ .
\label{eq:Ham-def}
\end{align}
Invariance with respect to time-translations implies that $\cH[u,u^*]$
is conserved by the flow generated by~\eqref{eq:NLS-hamform}. Additionally,
invariance under the transformation $u\mapsto e^{i\xi}u,\ \xi\in R^1$ implies
that 
\beq
\cN[u,u^*]\ =\ \int |u|^2\ dx =\ ||u||_2^2
\label{eq:N-def}
\eeq
is a conserved integral. \\

For the spatially translation invariant case, $\gamma=0$, NLS has the
Galilean invariance:
\beq
u(x,t)\ \mapsto\ u(x-vt,t)\ e^{i(xv -\frac{1}{2}v^2t)}, \ v\in R^1.
\label{eq:Galilei}
\eeq

\noindent{\it Well-posedness theory:}\ 
The functionals $\cH[\cdot]$ and $\cN[\cdot]$ are well defined on
$H^1(R^1)$, the space of functions $f$, for which $f$ and $\D_xf$ are
square integrable.  It is therefore natural to construct the flow for
initial data of class $H^1$.  In fact, it can be shown that, for
initial conditions $u_0=u(x,t=0)\in H^1(R^1)$, there exists a unique
global solution of NLS, $u\in C^0(R^1;H^1(R))$,  in the sense of the
equivalent integral equation:
\begin{subequations}
\begin{gather}
u(t)\ =\  U(t)u_0\ +i\int_0^t U(t-s)\  
 |u(s)|^2u(s)\ ds, \label{eq:NLS-ie}\\
U(t)\ \equiv\ 
 \exp\left(-iHt\right),\ H\equiv -\frac{1}{2}\D_x^2-\g\delta(x). 
 \label{eq:propagator}
\end{gather}
\end{subequations}
The spectral decomposition of $H$ is known explicitly~\cite{AGHH:88} and can
be used to construct $U(t)$ explicitly.

To show the existence of a solution to~\eqref{eq:NLS-ie} in $H^1$, we must
show the existence of a $C^0(R^1;H^1(R))$ fixed point of the mapping
$u(x,t)\mapsto J[u](x,t)$, given by the right hand side of~\eqref{eq:NLS-ie}.

We now outline the key ingredients of the proof.  To bound $J[u]$ and its
first derivative in $L^2$, we introduce the operator $\cA=I+P_cH$, where $P_c$
denotes the projection onto the continuous spectral part of $H$. Note that
$\cA$ is a nonnegative operator, since the continuous spectrum of $H$ is the
nonnegative real half-line.  Moreover, we expect $\|\cA^\frac{1}{2}f\|\ \sim
\|f\|_{H^1} \equiv\ \|(I-\D_x^2)^\frac{1}{2}f\|_{L^2}$. In fact, we shall use
that the following operators are bounded from $L^2$ to $L^2$: 
\beq
\cA^\frac{1}{2}(I-\D_x^2)^{-\frac{1}{2}},\
\cA^\frac{-1}{2}(I-\D_x^2)^\frac{1}{2} .  \nonumber 
\eeq 
This follows from the boundedness of the {\it wave operators} on 
$H^1$~\cite{Weder:99}.  Therefore, we have an equivalence of norms 
\beq C_1\ \| f\|_{H^1}\ \le\ \|
\cA^\frac{1}{2}f\|_{L^2}\ \le\ C_2\ \| f\|_{H^1} .
\label{eq:norm-equiv}
\eeq
Our formulation~\eqref{eq:NLS-ie} and introduction of $\cA$ is related
to the nice property that $\cA$, and hence also functions of $\cA$,
commute with the propagator $\exp(-iHt)$. We shall also use the
Sobolev inequality:
\beq 
|f(x)|^2 \le\  C\|f\|_{L^2}\|\D_xf\|_{L^2} ,
\label{eq:sobolev}
\eeq
and the Leibniz rule\ \cite{KP:88}:
\beq
\|(I-\D_x^2)^\frac{1}{2}(fg)\|\ \le 
\ C\left( \| f\|_{L^\infty}\ \|(I-\D_x^2)^\frac{1}{2}g\|_{L^2}\ +
 \ \|(I-\D_x^2)^\frac{1}{2}f\|_{L^2}\ \| g\|_{L^\infty} \right) .
\label{eq:leibniz}
\eeq

Since $U(t)$ is unitary in $L^2$, we have
\begin{align}
&\| \cA^\frac{1}{2} J[u]\|_{L^2}\ \nn\\
&\le\ \|\cA^\frac{1}{2} u_0\|_{L^2}\ +\ 
 \g\int_0^t \|  \cA^\frac{1}{2} |u(s)|^2u(s) \|_{L^2}\ ds\nn\\ 
&=\ \|\cA^\frac{1}{2} u_0\|_{L^2}\ +\ 
 \g\int_0^t \|  \left(\cA^\frac{1}{2}(I-\D_x^2)^{-\frac{1}{2}}\right)\cdot
 \left( (I-\D_x^2)^\frac{1}{2} |u(s)|^2u(s)\right) \|_{L^2}\ ds.
\label{eq:run1}
\end{align}
By~\eqref{eq:norm-equiv}),~\eqref{eq:sobolev}, and~\eqref{eq:leibniz},
\beq
\|J[u](t)\|_{H^1}\ \le\   
  C\| \cA J[u]\|_{L^2}\ \le\ C_1\|u_0\|_{H^1}\ +\ 
 C_2\ T\  
 \sup_{s\in [0,T]} \|u(s)\|_{H^1}^3 .
\label{eq:Jbound}
\eeq
Now assume that $u$ is such that $\sup_{s\in [0,T]}\|u(s)\|_{H^1} \le 2C_1$.
Then, by~\eqref{eq:Jbound}, by choosing $T<T_1$ sufficiently small.
 $\sup_{s\in [0,T]}\|J[u](s)\|_{H^1} \le 2C_1\|u_0\|_{H^1}$.
\beq
\|J[u](t)\|_{H^1}\ \le\ C_1\|u_0\|_{H^1}\ +\ C_2\ T\ \left( 2C_1\|u_0\|_{H^1}
\right)^3 .
\nonumber\eeq
It follows that for $0<T<T_1$ sufficiently small, the transformation
$J[\cdot]$ maps a ball $C^0([0,T];H^1(R))$ into itself. 
A similar calculation shows that 
\beq
\|J[u](t)\ -\ J[v](t)\|_{H^1}\ \le\  
 K\ T\ \left(2C_1\|u_0\|_{H^1}\right)^2\ 
 \sup_{s\in [0,T]}\ \| u(s)-v(s)\|_{H^1} ,  
\eeq
and therefore for $0<T<T_2\le T_1$, the transformation $J[\cdot]$
is a contraction on this ball. Therefore,  $J[\cdot]$ has a unique fixed point
in $C^0([0,T];H^1(R))$ for $T$ sufficiently small and local existence in time
of the flow follows.  Global existence in time follows from the {\it \'a
priori} bound on the $H^1$ norm of the solution implied by the
time-invariance of $L^2$ norm and Hamiltonian. 
\medskip

\noindent {\it Nonlinear bound states:}\ 
 Bound states are an important class of solutions having the form:
\beq u_b(x,t)\ =\ e^{-i\lambda t} \varphi(x;\lambda),\  \varphi\in L^2.
\label{eq:ub}
\eeq
For the linear Schr\"odinger equation, the functions $\varphi$ are  
eigenstates of a Schr\"odinger operator: $-\frac{1}{2}\D_x^2-\gamma\delta(x)$
and satisfy the equation
\beq
-\frac{1}{2}\varphi_{xx}\ -\ \g\delta(x)\varphi =
 \lambda\varphi .
\label{eq:linear-efn}
\eeq
Bound states are known to play a fundamental role in the general 
dynamics of the linear Schr\"odinger equation. This is a consequence of the
spectral decomposition of linear self-adjoint operators.

For NLS, such bound states satisfy the equation:
\beq
-\frac{1}{2}\varphi_{xx}\  -\ \left(\ \g\delta(x)\ +\ 
  |\varphi|^2\ \right)\varphi\  =\ 
 \lambda\varphi ,
\label{eq:varphi-eqn}
\eeq
and have the general character of
`nonlinear eigenstates',  although there is no rigorous decomposition
theory of solutions into such states, except in the  completely
integrable case $\g = 0$ \cite{ZS:72}. For this translation-invariant case
there is family of solitary traveling wave solutions~\eqref{eq:soliton}. 
These are Galilean boosts of the basic solitary
standing wave~\eqref{eq:soliton} with $v = 0$; see~\eqref{eq:Galilei}.
For $\g > 0$, the equation
is no longer translation-invariant and  we have the defect or
`pinned' states of~\eqref{eq:defectmode}. These two families
of nonlinear bound states are plotted in bifurcation diagram of
Figure~\ref{fig:L_v_freq}.  We note that for $\g=0$ the family of
solitons bifurcates from the zero state at zero frequency, the
endpoint of the continuous spectrum of the linearized operator
$-\D_x^2$ about the zero state.  For $\g>0$, the family of defect
states bifurcates from the zero state in the direction of the
eigenfunction of the linearized operator,  $-\D_x^2 - \g\delta(x)$, and
at the corresponding eigenfrequency  $\lambda = -\g^2/2$; see
\cite{RoseWeinstein:88} for a general discussion. \\

\noindent {\it Stability of nonlinear bound states:}\ 
An alternative characterization of the nonlinear bound states
$u_{\rm Sol}$ and $u_{\rm Def}$ is variational. The advantage of the 
variational characterization is that it can be used to establish
nonlinear stability  of the `ground state; 
see~\cite{Weinstein:86,RoseWeinstein:88}.
\begin{thm}
\label{thm:variational}

\noindent (I)\ The families of nonlinear bound state profiles 
$\eta\mapsto\varphi_{Sol}(x;\eta)$ for the case $\g=0$ and 
$a\mapsto\varphi_{Def}(x;a)$ for the case $\g>0$ can be characterized
variationally as {\it minimizers} of the Hamiltonian, $\cH$, subject to 
fixed $L^2$ norm, $\cN$:
\beq
\min_\varphi\  \cH[\varphi],\ \  \cN[\varphi]=\rho
\label{eq:var}
\eeq
Thus $u_{\rm Sol}$ and $u_{\rm Def}$ are called {\it ground states} in their
respective cases. Their associated frequencies, $\lambda(\rho)=-\eta^2/2$
and $\lambda(\rho)=-a^2/2$ arise as Lagrange multipliers for the constrained 
 variational problem~\eqref{eq:var}. As $\rho\to0$, $\lambda(\rho)\to0$,
 respectively, $\lambda(\rho)\to -\gamma^2/2$.

\noindent (II)\ Ground states are $H^1$ 
nonlinearly orbitally Lyapunov stable, {\it i.e.} if the initial data are 
$H^1$ close to a soliton (modulo the NLS symmetries), then 
the solution remains close to a soliton in this sense for all $t \in 
(-\infty,\infty)$. 
\end{thm}
\noindent For results on asymptotic stability of nonlinear ground states, 
see~\cite{SW:90,SW:92,PW:95,BP:93,Weder:00}.
\medskip

\noindent{\it High energy defect modes and solitons:}\ Solitons of the
translation-invariant NLS equation ($u_{\rm Sol}$ of~\eqref{eq:soliton}
for $\gamma=0$) may be related to the high energy ($L^2$ norm)
nonlinear bound states, $u_{\rm Def}$, in an appropriate limit. The
following summary is based on the variational principle of
Theorem~\ref{thm:variational}, and does not require the explicit
formula~\eqref{eq:defectmode} for the defect mode. This argument can
also be applied  to more general linear potentials $V(x)|u|^2$ in the
Hamiltonian in place of $\delta(x)|u|^2$, and to more general
nonlinearities.

Consider the variational problem~\eqref{eq:var}, in which we make
explicit the dependence of the Hamiltonian $\cH$ on  the defect
strength $\gamma$ by writing $\cH(u;\gamma)$ in place of the notation
of~\eqref{eq:Ham-def}.  Define $T_\rho[u](x) = \rho u(\rho x)$, for
$\rho>0$.  Then, $\frac{1}{\rho^3} \cH(T_\rho[u];\gamma) =
\cH(u;\frac{\gamma}{\rho})$.  If $\cI(\rho;\gamma)$ denotes the minimum
in~\eqref{eq:var}, we therefore have:
\begin{equation}
\frac{1}{\rho^3} \cI(\rho;\gamma) = \cI(1;\frac{\gamma}{\rho}) .
\nn \end{equation}
Note that, as $\rho\to\infty$, 
 $\cI(1;\frac{\gamma}{\rho})$ formally approaches $\cI(1;0)$,
the constrained minimum of the cubic NLS Hamiltonian $\cH(u;0)$,
 and that the extremizer is the classical one-soliton.
It  can be shown that  if
$u_{\rm Def}(x;a)$ denotes the nonlinear defect mode and
we define $U_D$ by $T_{1/a}[u_{\rm Def}](x;a) = U_D(x;a)$,
then $U_D(x;a)$ converges strongly to the
classical one-soliton of norm one. It follows that,
for large $a$, $u_{\rm Def}(x,a)$ looks more and more like a 
$T_a$-scaled soliton (a solitary standing wave) of the 
translation-invariant NLS.

\section{Direct Numerical Simulations of the PDE}
\label{sec:dns}

In this section we discuss simulations of the initial value problem
for the  NLS equation~\eqref{eq:NLS}.  All numerical experiments in
this section were performed using a modification of a finite
difference approximation due to Fei, P\'erez-Garc\'ia and
V\'azquez~\cite{FPV}, which conserves a discrete $L^2$ norm and, in
the absence of a defect, a discrete analog of the Hamiltonian.  The
method is accurate to second order in both space and time, and is
implicit only in its linear terms. Therefore, it requires a linear
equation to  be solved at each step. The Dirac delta function is
approximated either as a single point discontinuity, or by a smoother
function with very small support, with similar results in either case.

In the numerical experiments, a soliton is initialized far from the
defect with prescribed velocity $v$ and amplitude $\eta$ and is
allowed to propagate toward the defect location.  A wide variety of
behaviors is seen as parameters are varied.  For simplicity (and motivated 
by a scaling argument given in section~\ref{sec:ode}) we limit our study
to defects with strength $\gamma=1$. Therefore, the branch of nonlinear
 defect modes bifurcates at the frequency $-1/2$.  The nonlinear resonance
summarized in Figure~\ref{fig:L_v_freq} is useful in understanding the
various behaviors that are possible in this interaction.  In the
figures that follow, $|u|$ is plotted although the system's $L^2$
conservation might suggest plotting $|u|^2$.  This is done to render
visible the radiation that is shed during the interaction.

\begin{figure}
\begin{center}
\includegraphics[width=3in]{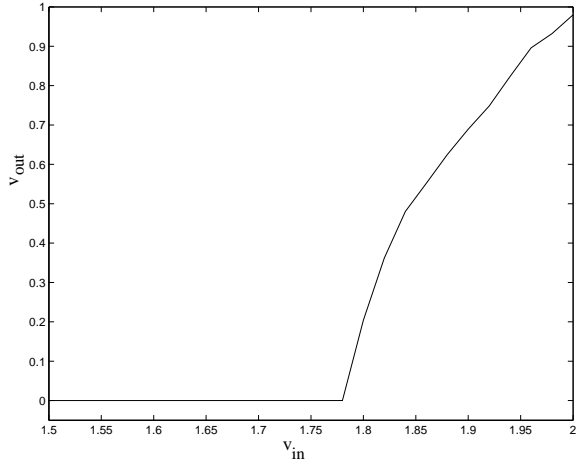}
\end{center}
\caption{Input versus output velocities of a soliton with $\eta=4$. }
\label{fig:V_in_out}
\end{figure}

\begin{figure}
\begin{center}
\includegraphics[width=2.25in]{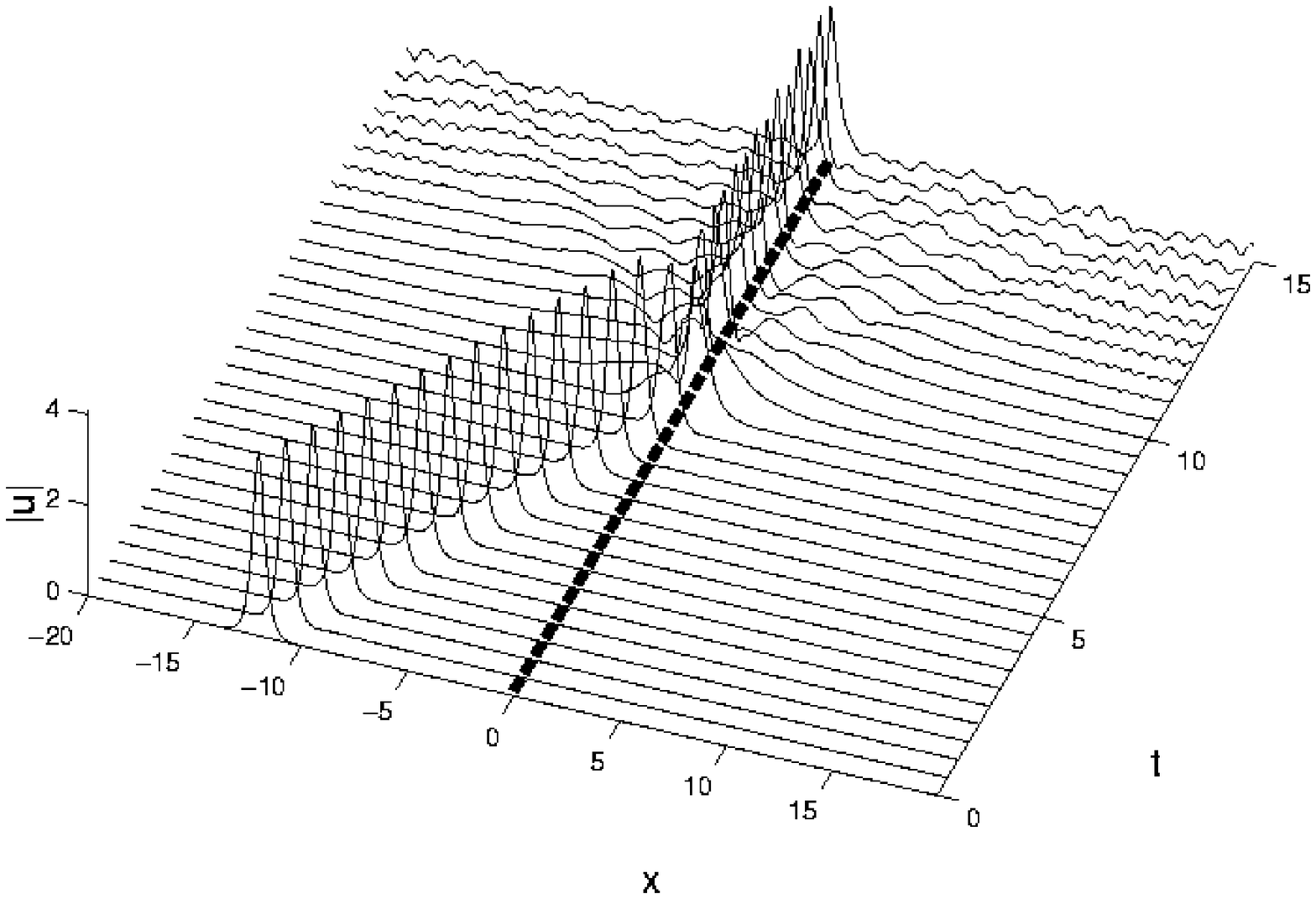}
\includegraphics[width=2.25in]{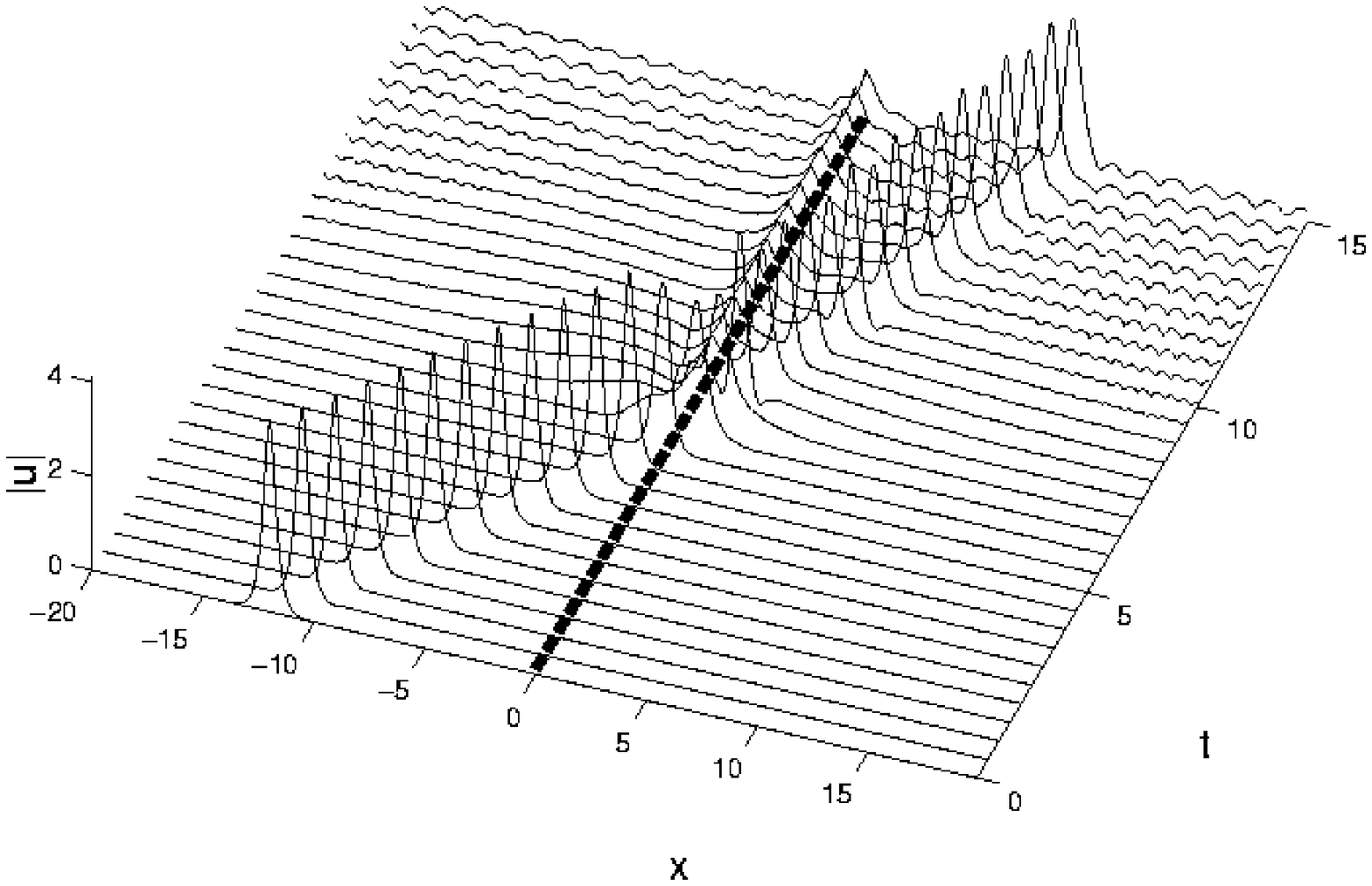}
\end{center}
\caption{The soliton amplitude $|u|$ after interaction with the
soliton, with large initial soliton amplitude $\eta=4$.  On the left,
a slower soliton is captured.  On the right, a faster soliton is
transmitted,  leaving behind a small defect mode. The transition from
capture to transmission is abrupt, occurring at a critical velocity
$v_c \approx 1.78$, as seen  in Figure~\ref{fig:V_in_out}. }
\label{fig:capt_trans}
\end{figure}

In the first set of runs, $\eta$ is set to 4, and $v$ is varied between
1 and 2. In this case, the solution remains mainly soliton-like, with
very little loss of energy to radiative modes.  There exists a
critical velocity, above  which the soliton is transmitted past the
defect, although with diminished speed, and below which the soliton
transfers its energy to a nonlinear defect
mode. Figure~\ref{fig:V_in_out} shows the input versus output
velocities, indicating the critical velocity $v_c \approx 1.78$.
Figure~\ref{fig:capt_trans} shows the evolution of $|u(x,t)|$ up to a time
somewhat after the interaction has taken place.  In the left hand figure,
the initial velocity was about $v=1.65$, and after some time the 
solution is a defect mode centered at $x=0$ with a small amount of
radiation. In the right hand figure, with initial velocity greater than
$v_{c}$, the soliton largely survives, although some of the
incoming soliton's  energy is captured by the defect and, as time
proceeds, eventually takes on the form of a small amplitude defect
mode.

\begin{figure}
\begin{center}
\includegraphics[width=2.25in]{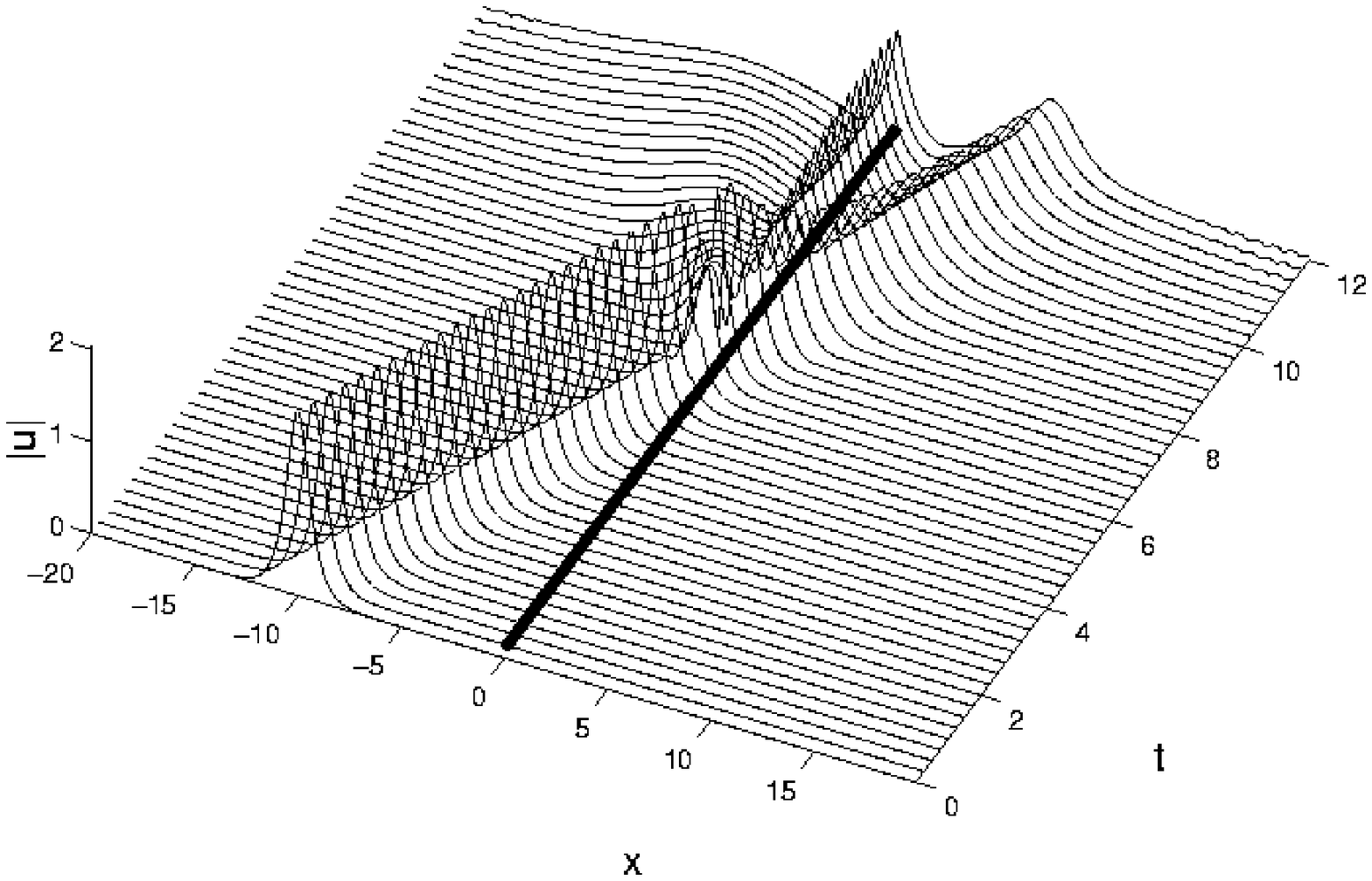}
\includegraphics[width=2.25in]{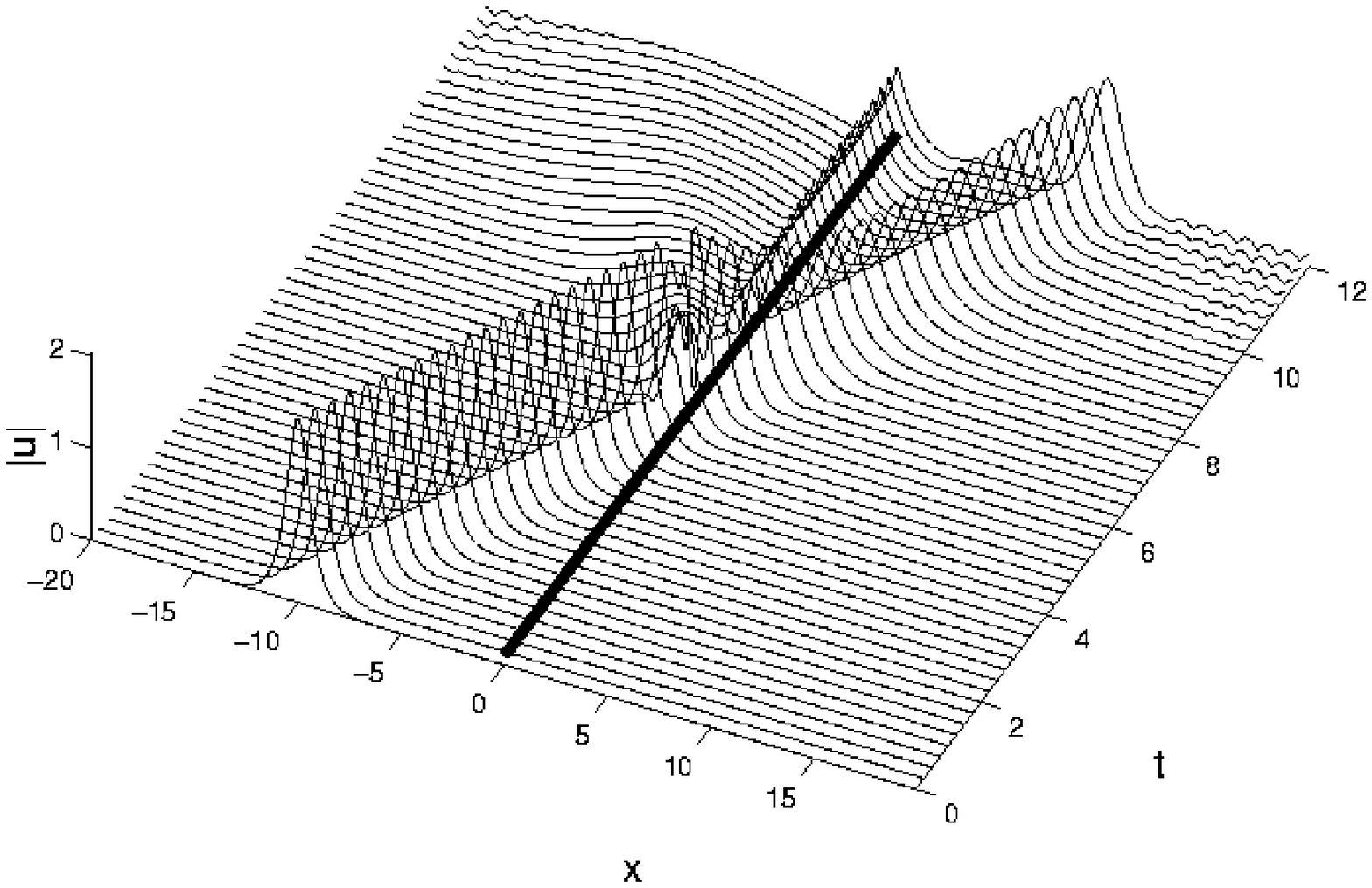}
\end{center}
\caption{A soliton with initial amplitude $\eta=2$ has a large amount of its
energy captured by the defect.  At left, a slower soliton with $v=1.5$ has a
substantial portion of its energy captured.  At right, as the incoming soliton's
velocity is increased to $v=1.75$ , less energy is captured and a larger
soliton gets through.}
\label{fig:eta_2}
\end{figure}

\begin{figure}
\begin{center}
\includegraphics[width=4in]{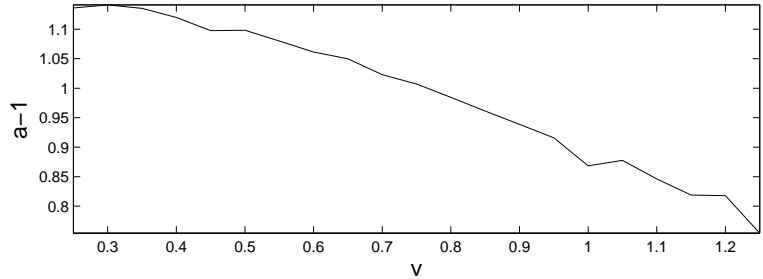}
\caption{For $\eta = 2$, as the input velocity is increased, the total
amount of energy captured by the defect mode decreases.}
\label{fig:captured_vs_v_in}
\end{center}
\end{figure}

When $\eta$ is reduced to 2, the behavior changes.  The
simulations show a more complicated nonlinear scattering process. The
pulse splits into three parts: reflected, captured, and transmitted.
In all cases, a significant defect mode is created. The faster the
soliton's initial speed, the smaller the defect mode remaining 
 in a neighborhood of
the origin and the larger the transmitted portion.  An example is shown in
Figure~\ref{fig:eta_2} and the phenomenon is summarized in
Figure~\ref{fig:captured_vs_v_in}, which shows how the fraction of energy
deposited into the defect mode decays monotonically as the input velocity
increases.

\begin{figure}
\begin{center}
\includegraphics[width=2.25in]{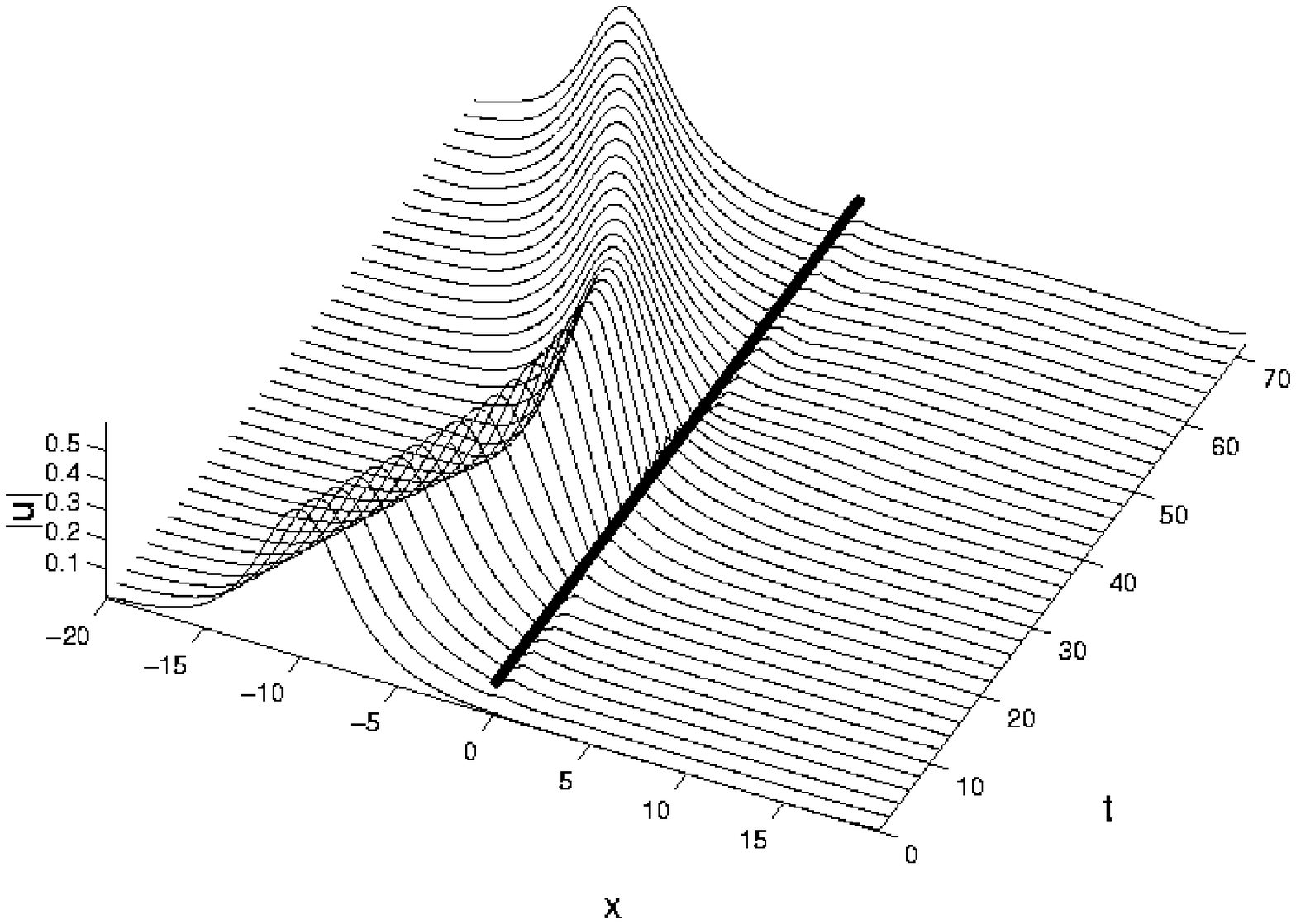}
\includegraphics[width=2.25in]{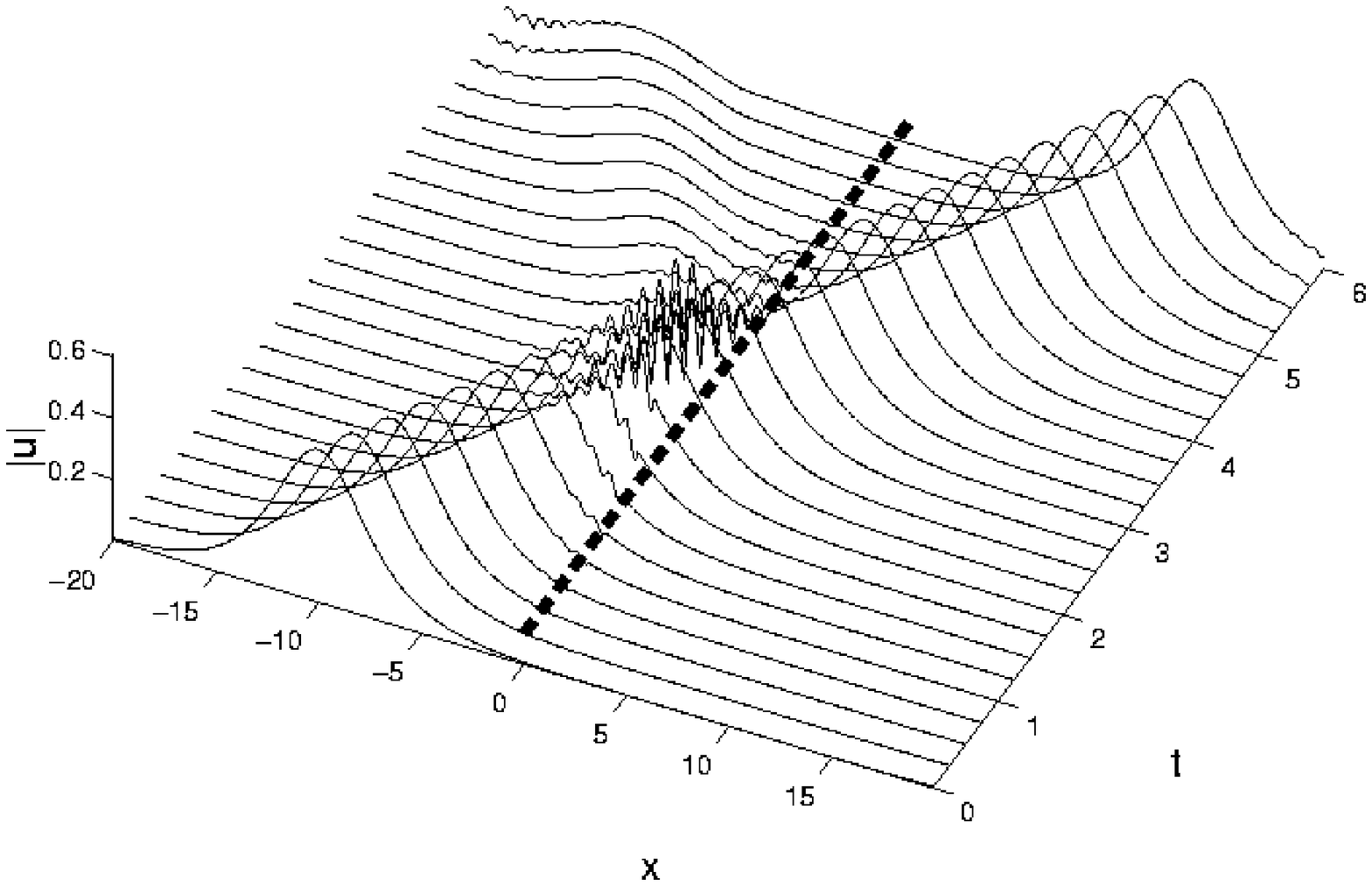}\\
\includegraphics[width=2.25in]{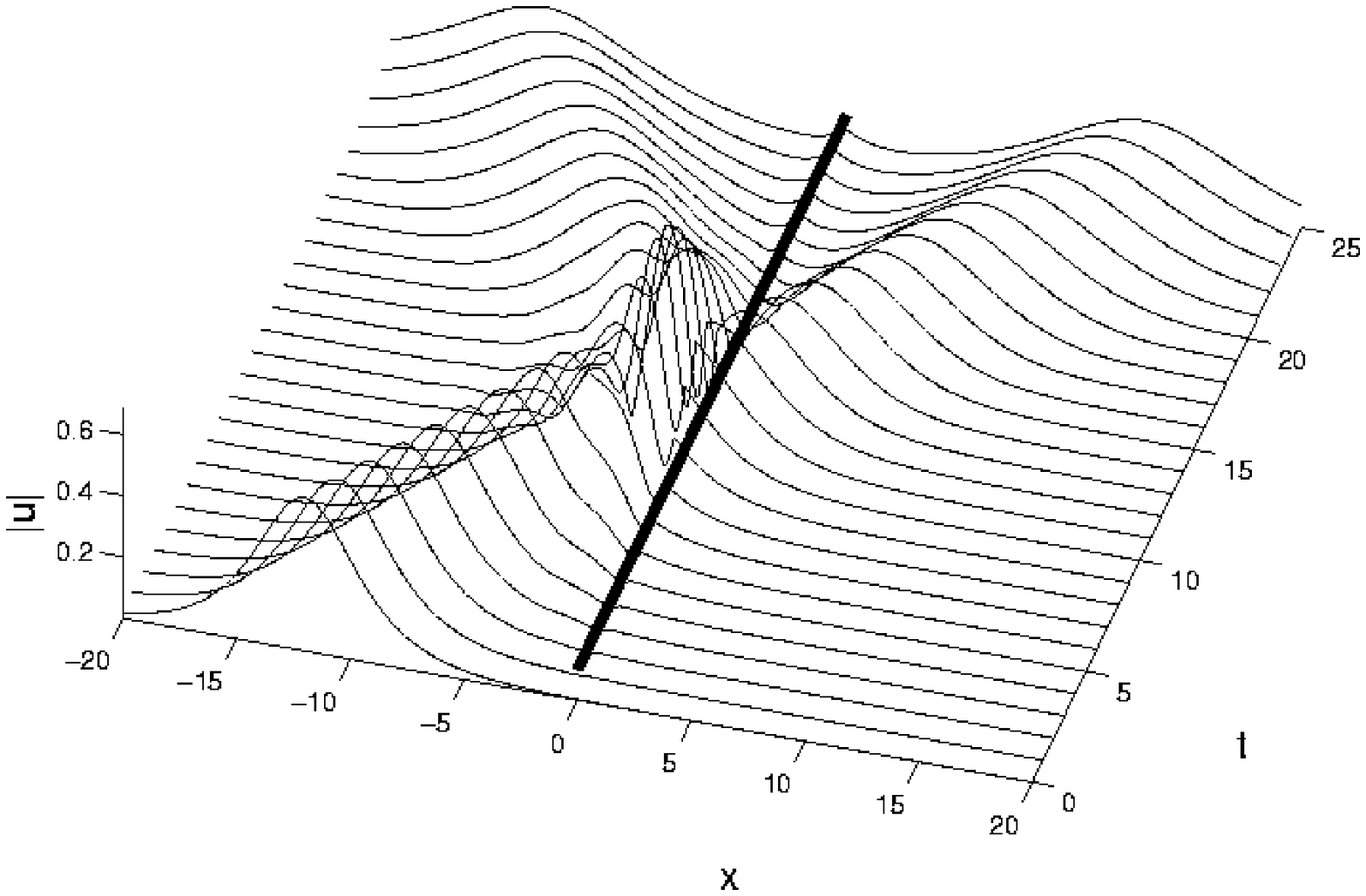}
\end{center}
\caption{When $\eta=0.5$, most of the wave is reflected with $v=0.25$
(left) or transmitted when $v=4$ (right). When $v=1$, the wave is
split nearly in half into reflected and transmitted portions.}
\label{fig:eta_half}
\end{figure}

When $\eta$ is further decreased to $0.5$, the behavior alters yet
again.  In this case, only a small amount of energy is captured by the
defect, while large amounts are reflected and transmitted; see
Figure~\ref{fig:eta_half}.  When the incoming velocity is sufficiently
small, the soliton appears to be completely reflected and when the
incoming velocity is sufficiently large, the soliton is nearly
completely transmitted. For intermediate velocities, the pulse is
split into a transmitted and a reflected wave. This nonlinear
scattering phenomenon is studied by Cao and Malomed~\cite{CM:95}, who
derive approximate reflection and transmission coefficients for the
interaction in the case of small $\eta$.

\subsection{Discussion of Simulations}

We wish to interpret the scattering of solitons by defects in terms of
the amplitude-frequency curves of Figure~\ref{fig:L_v_freq}. In the
previous section, as the soliton amplitude parameter $\eta$ is
decreased, two effects make the transfer of energy from the soliton to
the defect mode less efficient and enhance the scattering. At
first, for $\eta=2$, the amount of $L^2$ energy lost in the
interaction is increased, compared to the experiments with $\eta=4$.
Finally, for $\eta<\gamma$, there no longer exists a nonlinear defect
mode that resonates with the incoming soliton, and therefore almost no
energy is captured.

It was suggested in~\cite{GSW}, in the context of the nonlinear
coupled mode equations, that for sufficiently slow incident solitons,
a simple resonant energy transfer mechanism should hold. In particular, a
good approximate predictor of the distribution of trapped energy would
be given by the vertical projection of the point corresponding to the
incident soliton onto the corresponding point on the defect mode curve
with the same frequency. It turns out that this approximation is valid
in certain cases, but extensive further simulations have shown the 
general situation to be more complicated. In this subsection we 
explore this issue by means of the following auxiliary numerical 
experiment.

\begin{figure}
\begin{center}
\includegraphics[width=2.5in]{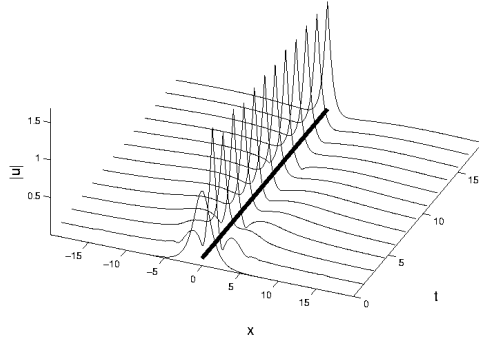}
\caption{A soliton initialized over a defect quickly decomposes into 
a defect mode plus radiation.}
\label{fig:even_solution}
\end{center}
\end{figure}

\begin{figure}
\begin{center}
\includegraphics[width=2.5in]{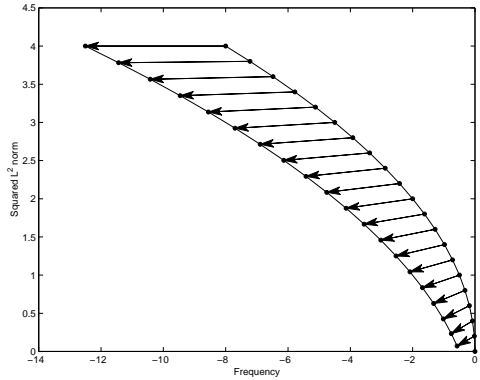}
\caption{Arrows connect soliton initial conditions to defect mode 
steady periodic solutions.}
\label{fig:capture_properties}
\end{center}
\end{figure}

We initialized a family of solitons with zero velocity and varying
amplitudes $\eta$, centered directly over the defect, and let them run
forward until they formed standing wave states (recall that the defect
standing wave~\eqref{eq:defectmode} is an exact solution of~\eqref{eq:NLS}). The solutions rapidly evolved into a combination of
nonlinear defect modes and radiation.  If the above picture applied,
then the defect modes thus produced should have almost the same
frequencies as the initial solitons. An example is shown in
Figure~\ref{fig:even_solution}, which was initialized with $\eta=1$
and $\gamma=1$.  Figure~\ref{fig:capture_properties} reproduces the
two curves of Figure~\ref{fig:L_v_freq} with arrows connecting each
soliton's initial conditions to the periodic solution of the
corresponding defect mode following the interaction. The arrows are
far from vertical, and show a consistent downward frequency
shift. Evidently, in this experiment the above `direct' resonant
exchange mechanism does not apply.

However, this finding does not contradict the role of resonant energy transfer
in trapping an incoming soliton.  In the above experiment, in which the
`incident' soliton has velocity $v=0$, the mechanism differs from that
involving a moving soliton. Were the nonlinear term absent, the results of
this auxiliary experiment would be well-understood.  Spectral theory dictates
that the initial condition decomposes into a bound state and radiative modes,
and there is no role in this interaction for the soliton's internal
frequency. In the presence of nonlinearity, the behavior is essentially the
same, although we can no longer compute the captured solution as the
projection of the initial conditions onto the bound state. The frequency is
shifted because the solution is in some sense finding the nonlinear projection
of the initial condition onto the family of nonlinear defect
modes~\eqref{eq:defectmode}, which has only a slightly smaller $L^{2}$ norm.
As $a$ is increased, the energy ($L^{2}$ norm) is increased and, as discussed
at the end of section~\ref{sec:overview}, the ground state defect mode
approaches a scaled one-soliton centered at $x=0$.

By contrast, {\it in the numerical simulations of moving solitons, the
defect mode is initially forced by the tail at the soliton's leading
edge, for which the soliton's internal frequency is important.} The
defect mode may grow by resonantly extracting energy from the tail,
before the bulk of the soliton even reaches the defect. Consequently,
when the soliton reaches the defect, the defect mode is large enough
that the two modes may interact, and energy may flow from one mode to
the other. In the next section, we introduce an ordinary differential
equation model of this interaction.

\section{A model of soliton-defect interactions}
\label{sec:ode}

Forinash, Peyrard, and Malomed~\cite{FPM:94} studied the interaction
of a soliton with a linear defect mode using a collective coordinate
ansatz to derive a set of approximate equations to describe the
evolution of a finite set of variables that characterize the two
modes.  Their model yielded a complicated set of
differential-algebraic equations which was difficult to understand
analytically.  Here we modify and slightly simplify their ansatz. In
particular, we approximate the solution, $u$, as the sum of a
time-modulated soliton and a time-modulated {\em nonlinear\/} defect
state:
\beq
u\ =
\ u_{\rm{S}}(x;\eta,Z,V,\phi)  + u_{\rm{D}}(x;a,\phi,\psi),
\label{eq:ansatz}
\eeq
where $u_{\rm{S}}$ is a generalized soliton
of the form
\beq
u_{\rm{S}} = \eta\sech(\eta x-Z) \; e^{i(Vx-\phi)} ,
\label{eq:u_S}
\eeq
and $u_{\rm{D}}$ is a generalized bound state of the form
\beq
u_{\rm{D}} = a\sech{\left( a \abs{x} + \tanh^{-1}\frac{\g}{a} \right)}
 \; e^{-i(\phi + \psi)}.
\label{eq:u_B}
\eeq

In~(\ref{eq:u_S},~\ref{eq:u_B}), the variables $\eta$, $Z$, $V$,
$\phi$, $a$, and $\psi$ are all allowed to depend on $t$. Note that
$\norm{u_S}^2_{L^2}=2\eta$ and $\norm{u_D}^2_{L^2}=2(a-\gamma)$. The
Lagrangian functional of  the NLS equation is evaluated at $u$ (by
integrating over the spatial domain), and the resulting function is
then interpreted as an effective finite-dimensional Lagrangian whose
Euler-Lagrange equations determine the evolution of these
variables. Instead of seeking stationary configuations of 
the (true) Lagrangian with respect
to admissible variations of the field variable $u(x,t)$, we
consider only variations of the type allowed by the
time-dependent variables of the ansatz. This technique, often known as the 
collective coordinate or variational method, has a long history and is well
summarized in a recent survey by Malomed~\cite{M:02}.

The NLS Lagrangian (cf.~\eqref{eq:Ham-def}) is
$$
\cL = \intinf \frac{i}{2}(u^*u_t- u u^*_t) - \frac{1}{2}\abs{u_x}^2 + \frac{1}{2}\abs{u}^4
+\g \d(x) \abs{u}^2 dx .
$$
We evaluate this integral for our ansatz
(\ref{eq:ansatz}-\ref{eq:u_B}). Unlike~\cite{FPM:94} and most other
analyses of this type of which we are aware, we do {\em not\/} assume
that $a$ is small, and we include high-order terms involving $a$. We
do make the simplifying assumption that all interaction between the
modes $u_{\rm{S}}$ and $u_{\rm{D}}$ takes place through terms
involving the delta function.  This is, in part, justified because the
other interaction terms involve oscillatory integrals which will
average out to be much smaller than the terms retained.\footnote{Here we are 
essentially anticipating that this effective Lagrangian is a normal form
to which the `exact' one is equivalent up to change of variables. 
The assumption that the integrals are oscillatory is
violated when $|V|\ll 1$, i.e. when the soliton is stalled.  The
assumption that the overlap integral is small will be violated when
$|Z|\ll 1$, i.e. when the soliton is close to the defect.}  The
resulting effective Lagrangian is given by
\begin{multline}
\cL_{\rm eff} = 2\eta\dot\phi - 2 Z\dot V + 2(a-\g)(\dot \phi + \dot\psi) 
+ \frac{1}{3}\eta^3 - \eta V^2 
+\frac{1}{3} a^3 +  \g \eta^2 \sech^2{Z}\\ 
+ 2\g\eta \sqrt{a^2-\g^2}\sech{Z}\cos{\psi} .
\label{eq:Lagrangian}
\end{multline}

This Lagrangian has an associated Hamiltonian, conserved by the 
Euler-Lagrange (and Hamilton's) equations:
\beq
 H = - \frac{1}{3}\eta^3 + \eta V^2
-\frac{1}{3} a^3 -  \g \eta^2 \sech^2{Z}
- 2\g\eta \sqrt{a^2-\g^2}\sech{Z}\cos{\psi}.
\label{eq:Hamiltonian}
\eeq
Since the Hamiltonian is independent of $\phi$, its conjugate momentum
\beq
p_{\phi}=\pdiff{\cL_{\rm eff}}{\dot \phi} = 2\eta + 2(a-\g)
\label{eq:p-phi}
\eeq
is conserved by Noether's theorem~\cite{Arn:78}. Notice that $p_{\phi}
= \norm{u_S}^2_{L^2}+ \norm{u_D}^2_{L^2}$. This conservation law is 
the analogue of the $L^2$ conservation law for NLS,~\eqref{eq:N-def}. 
The phase space of this three degree-of-freedom system can be
therefore be expressed as the cross product of the reduced
4-dimensional $(Z,V,a,\psi)$-phase space and the 2-dimensional
$(\eta,\phi)$-phase space, with trivial dynamics on the latter,
determined from the reduced system via~\eqref{eq:p-phi} and a
quadrature (cf.~(\ref{eq:ode}-\ref{eq:phi}) below). In the analysis that
follows, we may therefore regard the constant of motion
\beq
c = \eta(t) + a(t) \ \ ( \ge \g)
\label{eq:c}
\eeq
as a parameter determined by the initial conditions, and study
evolution on the reduced $(Z,V,\psi,a)$-space. Here and
henceforth $c$ denotes this constant, and should not be confused with its
conventional usage to denote the speed of light.

We note further that the resulting equations are invariant under the
rescaling $a \mapsto \g a, \; c \mapsto \g c, \; t \mapsto t/\g^2$: implying
that, without loss of generality, the parameter $\g$ may be set to equal to
one.  This symmetry belongs only to the reduced system of ODEs, not to
the original NLS equation~\eqref{eq:NLS}, but motivates our decision to 
perform simulations only with $\gamma=1$.
Then $c$ is the only parameter remaining in the evolution equations, and
these equations are in fact canonically Hamiltonian for the `scaled'
Hamiltonian $H/2$, with $H$ of~\eqref{eq:Hamiltonian} written in the
Lagrangian coordinates:
\beq
H = -\frac{c^3}{3} + (c-a) 
 \left( ca + V^2 - (c-a)  \sech^2 Z \right) 
 - 2 (c-a) \sqrt{a^2-1} \sech Z \cos \psi.
\label{eq:H1}
\eeq
The final reduced equations, to be studied in 
Section~\ref{sec:ode_analysis} below, are now:
\begin{subequations}
\begin{align}
\dot Z &= (c-a)V,  \label{eq:odeZ}\\
\dot V &= -(c-a)^2 \sech^2{Z} \tanh{Z} -
	  (c-a)\sqrt{a^2 - 1} \sech{Z} \tanh{Z} \cos{\psi} , \\ 
\dot \psi &= \frac{c^2 - 2ca - V^2}{2} + (c-a) \sech^2{Z} +
\frac{(2a^2 - ca - 1)}{\sqrt{a^2 - 1}}\sech{Z} \cos{\psi} 
\label{eq:psidot} , \\
\dot a &= -(c-a)\sqrt{a^2 - 1}\sech{Z} \sin{\psi},
\end{align}
\label{eq:ode}
\end{subequations}
with $\phi$ evolving according to:
\beq
\dot \phi = - \frac{(c-a)^2 - V^2}{2} - (c-a)\sech^2{Z} - \sqrt{a^2 - 1}
\sech{Z} \cos{\psi} .
\label{eq:phi}
\eeq

Before analyzing these ODEs, we describe numerical experiments that
reveal interesting  interactions between the soliton and the defect
mode, and suggest specific questions.

\section{ODE Simulations of the initial value problem}
\label{sec:odesim}

We now describe a set of numerical experiments for the ODE~\eqref{eq:ode}
analogous to those discussed in Section~\ref{sec:dns} for the PDE. We
initialize a soliton at $Z(0) =-Z_0$ with $|Z_0| \gg 1$, set equal to 20 for
these simulations.  The velocity parameter is set to $V(0)=V_{\rm in}$,
propagating rightward toward the defect. For subsequent comparison with the
PDE simulations, we note that the variable $V$ in~\eqref{eq:ode} is related to
the soliton velocity $v$ of~\eqref{eq:soliton} via $v = \eta V$
by~\eqref{eq:odeZ} and~\eqref{eq:c}.  The soliton amplitude is set to
$\eta(0)=\eta_0$, and we assume there is no energy initially in the defect
mode, so $a(0) = \gamma = 1$.  Thus the relation~\eqref{eq:c} fixes the
constant $c$. (Because of the singularity of Equation~\eqref{eq:psidot}, we
set $a(0) = 1 + \varepsilon$, where $\varepsilon \ll 1$; values between
$\varepsilon = 10^{-8}$ and $\varepsilon=10^{-5}$ were used in these
computations.  For some of the effects seen, $\varepsilon$ needs to be set to
the small end of this range to get stable results, due to the singularity
in~\eqref{eq:psidot}. )  Finally, we set $\psi(0) = 0$. Note that, for large
$|Z|$, the $\psi$ dependence becomes exponentially weak, so we do not expect
strong $\psi$-dependent effects.

We choose a representative set of $\eta(0)$, or equivalently, $c$ values.  For
each fixed $\eta(0)$ we allow $V_{\rm in}$ to vary over a range of values, and
numerically integrate~\eqref{eq:ode} until the soliton center, $Z(t)$, reaches
the defect ($Z(t)\sim0$), (eventually) exits the defect region, and reaches $Z
= +Z_0$ or $Z = -Z_0$ at, say $t=T$. By a variant of the Poincar\'e recurrence
theorem, as in~\cite{GHW:01}, we can show that the soliton must eventually
escape any bounded set containing the defect; cf. Section~\ref{sec:P_inf}
below.  We then plot the soliton's outgoing velocity parameter $V_{\rm out}$
(related to the physical velocity by~\eqref{eq:odeZ}), and the amplitude
$(a(T)-1)$ of the defect mode as functions of $V_{\rm in}$.  For $Z_0$
sufficiently large and $\varepsilon$ sufficiently small, the initial value of
the phase difference $\psi$ was indeed found to be unimportant in determining
the values of $V(T)$ and $(a(T)-1)$.  In interpreting these results, it is
important to realize that the dynamics takes place in four-dimensional phase
space, and that the figures merely show projections of trajectories on lower
dimensional subspaces.

\subsection{Experiment 1: large $\eta$.}
\label{sec:bigeta}

\begin{figure}
\begin{center}
\includegraphics[width=3.5in]{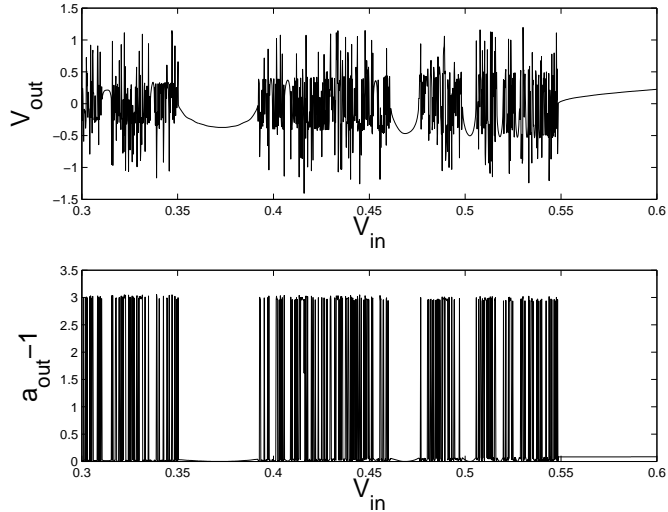}
\end{center}
\caption{(Top) Outgoing velocity vs. incoming velocity of solitons
incident on defect via ODE simulation with $\eta=4$. (Bottom)
Amplitude $a-1$ of nonlinear defect mode after passing of the
soliton.}
\label{fig:in_out_eta_4}
\end{figure}

We observe several distinct types of behavior. The behavior for
$\eta=4$ is shown in Figure~\ref{fig:in_out_eta_4}. In this case we
find a sharp change in behavior at a critical velocity parameter $V_c
\approx 0.55$.  Above this velocity, solitons pass through the defect
without significant interaction, merely decreasing their velocities
and transferring a little energy to the defect mode.  Below $V_c$,
however, the soliton interacts with the defect, oscillating within the
defect region a finite and apparently random number of times before
being ejected either to the right or the left. It is also striking
that the amount of energy remaining in the defect mode seems to be
restricted to two levels: either $a-1 \approx 3$ or $a-1 \ll 1$. This
behavior is apparently governed by the structure of the  invariant
manifolds of degenerate fixed points at $|Z| = \infty$. Since there
exist solutions to~\eqref{eq:ode} with $Z$ bounded, as well as
solutions which approach $Z=\infty$ with $V >0$, these must be
separated by solutions for which $Z \to \infty$ while $V \to 0$. Along
with the apparantly arbitrarily fine structure of transmission and
reflection zones, Figure~\ref{fig:in_out_eta_4} further shows wide
reflection windows of the type reported in many previous studies,
e.g.~\cite{AOM:91,CSW:83,FKV:92,YT:01}. 

\begin{figure}
\begin{center}
\includegraphics[width=3.5in]{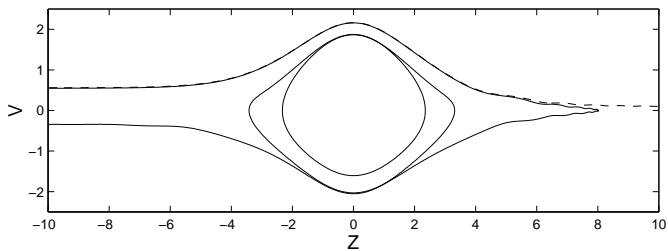}
\end{center}
\caption{$(Z,V)$ phase space plots of captured and transmitted 
trajectories just above (dashed) and below (solid) critical velocity. }
\label{fig:captured_transmitted}
\end{figure}

In Section~\ref{sec:P_inf}
we will investigate the stable and unstable manifolds that are
responsible for this behavior, but we give a brief preview here.
Figure~\ref{fig:captured_transmitted} shows the $(Z,V)$ projections of
trajectories with nearby initial velocities on either side of $V_c$
($V(0) = V_c \pm .002$). The solid curve comes close to an orbit
apparently asymptotic to $(Z, V) = (\infty, 0)$, and then turns back
and is transiently captured before eventual reflection, while the
dashed curve approaches infinity with $V$ bounded above zero, and is
transmitted without further interaction. This `separatrix' behavior is
repeated in the multiple transmission and reflection windows for $V <
V_c$ (Figure~\ref{fig:in_out_eta_4}) and is reminiscent of that found
in our earlier study of an ODE model of kink-defect interaction in the
sine-Gordon equation~\cite{GHW:01}, and shown there to be related to
the homoclinic tangle formed by transverse intersection of stable and
unstable manifolds of periodic orbits at $|Z| = \infty$: compare
Figure~\ref{fig:in_out_eta_4} with Figure~3.2 of~\cite{GHW:01}. 

The ODEs' behavior should be compared with the direct numerical simulations
of solitons with $\eta=4$, as shown in Figure~\ref{fig:V_in_out}. The
critical velocity is overestimated by $24 \%$ ($v_c = \eta V_c = 2.21$
cf. $1.78$ for the PDE), and the PDE simulations show no evidence of
the fine structure of transmission and reflection zones below $v_c$.
This difference, partially due to neglect of radiation damping in the ODEs,
is discussed in section~\ref{sec:summary}.

\subsection{Experiment 2: medium $\eta$.}
\label{sec:modeta}

\begin{figure}
\begin{center}
\includegraphics[width=3.5in]{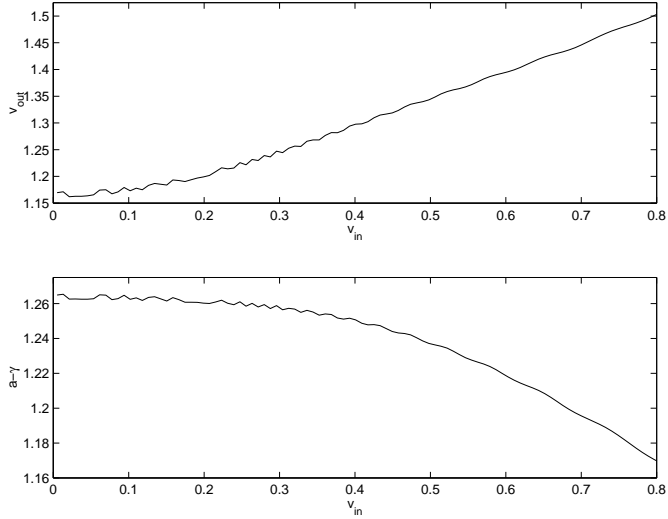}
\end{center}
\caption{(Top) Outgoing velocity parameter V vs. incoming V of solitons
incident on defect via ODE simulation with $\eta=2$. (Bottom)
Amplitude $a-1$ of nonlinear defect mode after passing of the soliton.}
\label{fig:in_out_eta_2}
\end{figure}

In Figure~\ref{fig:in_out_eta_2}, computed for $\eta=2$, we see a
quite different picture. Here and in the third numerical experiment,
below, $-\eta^2/2 < -\gamma^2/2 = -1/2$ and resonant interactions can
and do take place (cf. Figure~\ref{fig:L_v_freq}).  In this case there
is no transition between transmission and reflection: the soliton
travels monotonically rightward and is {\em always}  transmitted
without transient capture or oscillations about the defect.  More
strikingly, the output velocity appears to approach a finite limit
$V_{\rm out} \approx 1.17$ as $V_{\rm in} \to 0$, while the amount of
energy captured approaches $a-1 \approx 1.26$ (all the soliton's 
 energy
would be captured if $a-1 =2$).  As the initial velocity increases, so
does the output velocity, while the energy
  transferred from the soliton
to the defect mode decreases. This is not surprising, since the
duration of the interaction decreases with increasing soliton speed.

\subsection{Experiment 3: small $\eta$.}
\label{sec:smalleta}

\begin{figure}
\begin{center}
\includegraphics[width=4in]{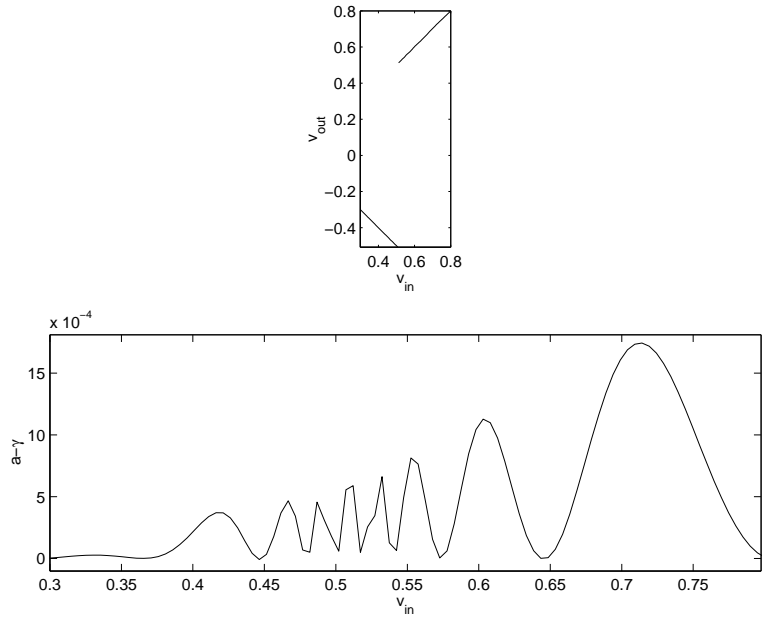}
\end{center}
\caption{(Top) Outgoing velocity vs. incoming velocity of solitons
incident on the defect via ODE simulation with
$\eta=\frac{1}{2}$. (Bottom) Amplitude $a-1$ of nonlinear defect mode
after passing of the soliton; note vertical scale is ${\mathcal{O}}(10^{-3})$.}
\label{fig:in_out_eta_half}
\end{figure}

For Figure~\ref{fig:in_out_eta_half}, we set $\eta = 0.5 < \g$; in
this case the soliton is reflected if $V_{\rm in}$ lies below a critical
velocity $V_c \approx 0.51$ and transmitted if $V_{\rm in} > V_c$.  We
also see that the defect mode has a final amplitude of order
$10^{-3}$, absorbing little of the soliton's energy. We note that
$-\gamma^2/2 = -1/2 < -\eta^2/2$ corresponds to the region in
Figure~\ref{fig:L_v_freq} in which the soliton has no resonant defect
mode `partner' with the same temporal frequency, and hence that
appreciable interactions are unlikely~\cite{GSW}.

\begin{figure}
\begin{center}
\includegraphics[width=3.5in]{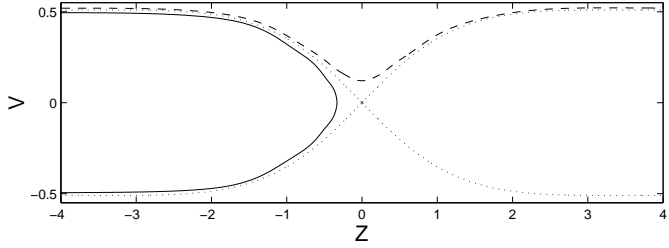}
\end{center}
\caption{Phase space projections of trajectories above (solid) and
below $V_c$ (dashed), showing strong evidence for a saddle point.
Projections of stable and unstable manifolds of the saddle on ${\cP_0}$
are shown as dotted curves.}
\label{fig:reflect_transmit}
\end{figure}

Figure~\ref{fig:reflect_transmit} shows evidence that the solution
passes near $Z = V = 0$, presumably approaching and leaving the
neighborhood of a hyperbolic invariant set on this subspace. In
Section~\ref{sec:ode_analysis} we shall show that this subspace
indeed contains a fixed point of saddle-center type surrounded 
by a family of periodic orbits whose stable manifolds serve as
separatrices. In Figure~\ref{fig:reflect_transmit} we also show
projections onto $(Z,V)$-space of the numerically determined stable
and unstable manifolds of the fixed point, along with two
trajectories: one with asymptotic velocity larger than the limiting
value on the unstable manifold, which is transmitted, and one with
asymptotic velocity smaller than the limiting velocity, which is
reflected.

\section{Analysis of ordinary differential equations}
\label{sec:ode_analysis}

The numerical experiments described above reveal two broad types of
behavior.  For small values of $\eta$, the soliton traveling near the
critical velocity appears to approach a hyperbolic fixed point or
periodic orbit on $(Z, V) = (0, 0)$: see
Figure~\ref{fig:reflect_transmit}. For larger values of $\eta$, it
oscillates about $(Z, V) = (0, 0)$ as if around an elliptic fixed
point, and also follows orbits asymptotic to $(Z, V) = (\pm \infty,
0)$ before turning: see Figure~\ref{fig:captured_transmitted}. To
interpret these observations we now analyze the ODE system~\eqref{eq:ode}, seeking a (partial) understanding of its global phase
space structure.

We first note that the sets $a=c$ and $a=1$ are invariant for the flow
(although the vector field is singular on the latter), and bound the
physically admissible region. When $a=1$ all the energy resides in the
soliton; when $a=c$ it all resides in the defect mode.  Letting ${\cI}
= [1,c]$ denote the closed interval, the phase space of
Equation~\eqref{eq:ode} is  $(Z,V,\psi,a) \in \RR^2 \times S^1 \times
{\cI}$. We also note the following (reversibility) symmetry group under
which~\eqref{eq:ode} is equivariant:
\begin{subequations}
\begin{align}
(Z, V, \psi, a, t) & \rightarrow (-Z, V, 2 \pi - \psi, a, -t) , \\
(Z, V, \psi, a, t) & \rightarrow (Z, -V, 2 \pi - \psi, a, -t) .
\end{align}
\label{eq:symm}
\end{subequations}
We shall use this below.

There is a family of solutions at $Z= \pm \infty$ with $V$, $a$, and
$\dot \psi$ constant, which correspond to the uncoupled propagation
and oscillation of the two modes when the soliton is infinitely far
from the defect. The subset of these solutions with $V=0$ form a
degenerate family of periodic orbits `at infinity,' parameterized
by $a = a_{\infty}$ and filling the annulus (or finite cylinder)
\beq
 {\cP_{\infty}} = \{ (\psi,a) | V = 0, \; |Z| = \infty \}.
\label{eq:Pinfty}
\eeq
We note
that by~\eqref{eq:psidot} and~\eqref{eq:c}, 
 $\dot{\psi} = (\eta^2 - a_{\infty}^2)/2$  on
${\cP_{\infty}}$, so that the frequency of these orbits is nonzero
provided $a_{\infty} \neq \eta$ or, equivalently, $a_{\infty} \neq
c/2$. As in~\cite{GHW:01} we may employ a transformation of the form
$q = \sech Z, \; p = V$ to bring these orbits to the origin in
$(p,q)$-space, and then apply McGehee's stable manifold
theorem~\cite{M:73} to prove the existence of invariant manifolds for
the fixed point $(p,q) = (0,0)$ in an appropriate (local) Poincar\'{e}
map. This shows that each periodic orbit in ${\cP_{\infty}}$ has
two-dimensional stable and unstable manifolds, so that
$W^s({\cP_{\infty}})$ -- the stable manifold of ${\cP_{\infty}}$
itself -- is three dimensional and hence locally separates the
four-dimensional phase space. Indeed, $W^s({\cP_{\infty}})$
separates orbits that escape to infinity (transmitted solitons) from
those that are reflected to interact with the defect mode again.

Studying the analogous sine-Gordon kink-trapping problem
in~\cite{GHW:01}, we used isoenergetic reduction and Melnikov's
method~\cite{M:63,GH:83} to prove that the stable and unstable
manifolds of each periodic orbit, restricted to their common energy
manifold, intersect transversely, and hence that Smale horseshoes
exist~\cite{GH:83}. We then appealed to phase space transport
theory~\cite{RKW:90,W:92} to unravel the structure of sets of initial
data that are transiently captured before eventually being transmitted
or reflected. We proceed in the same manner in
Section~\ref{sec:P_inf}, although we have to introduce an artificial
small parameter, and the set of solutions to which the standard
reduction procedure applies is limited, since it requires that the
frequency $\dot{\psi}$ not change sign (in the process one replaces
time by $\psi$), and this holds only for large $c$, depending on $a$;
cf. Equation~\eqref{eq:psidot}. In particular, it does {\em not\/}
hold for many physically relevant parameter values, including those
corresponding to initial data with (almost) all the energy in the
soliton $(a \approx 1)$. Nonetheless, the analysis does provide  some
understanding of the large $\eta$ simulations of
Section~\ref{sec:bigeta}.

There are further invariant manifolds that play an
important r\^{o}le in the fate of solutions. They belong to orbits on
a second annulus 
\beq
{\cP_0} = \{ (\psi,a) | V = 0, \; Z = 0 \}
\label{eq:P0}
\eeq
that is also invariant under the flow, and on which the ODEs are
integrable. Solutions on ${\cP_0}$ correspond to a soliton stalled
over the defect and periodically exchanging energy with the defect
mode. The orbit structure on ${\cP_0}$ depends upon $c$ and may be
derived from the level sets of the Hamiltonian function $H$ restricted
to ${\cP_0}$:
\beq
H_{\cP_0}(\psi,a) = -\frac{c^3}{3} + (c-a)
\left( ca - (c-a) - 2 \sqrt{a^2-1} \cos \psi \right) .
\label{eq:HP0}
\eeq

\begin{figure}
\begin{center}
\includegraphics[width=2.7in]{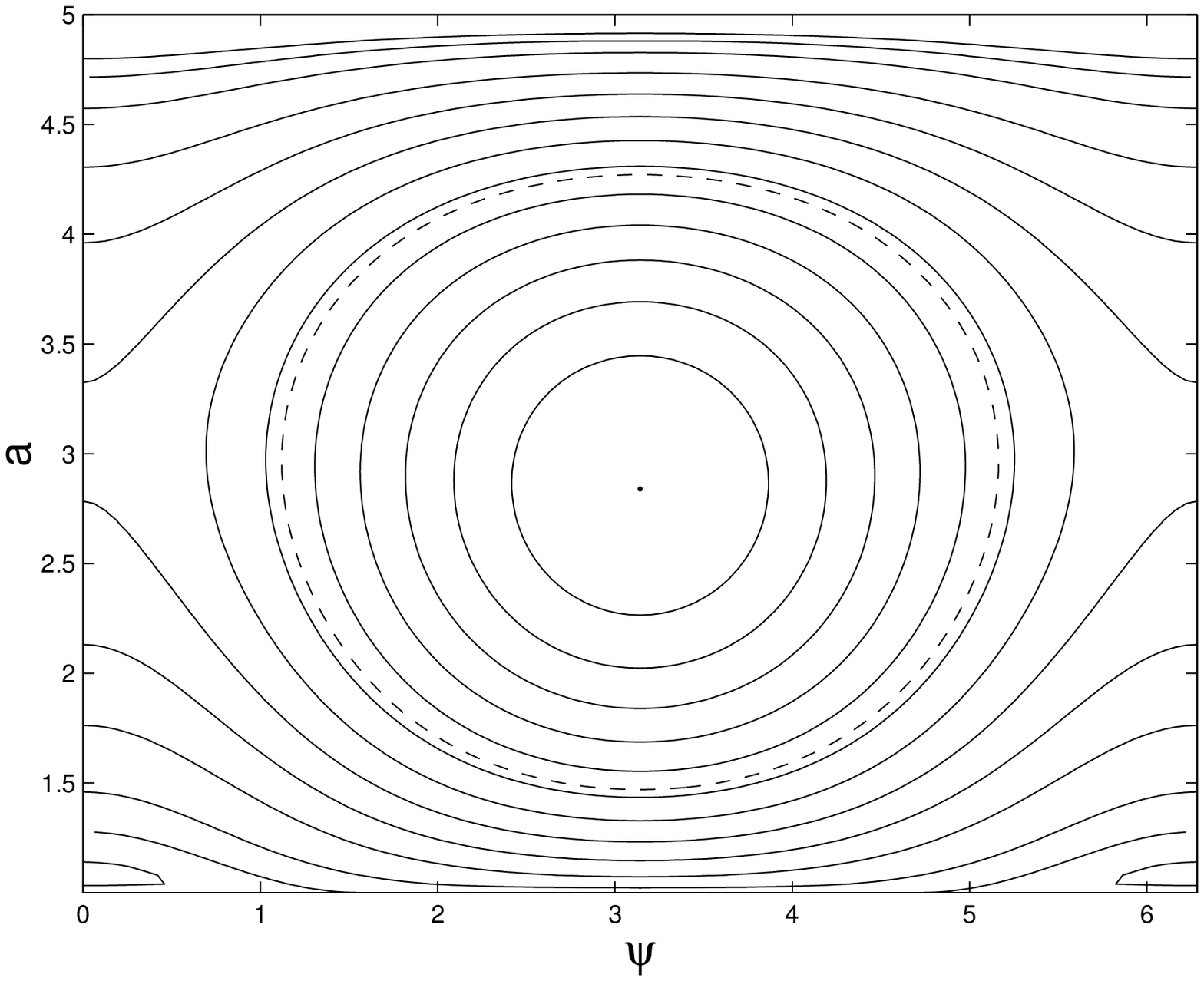}
\includegraphics[width=2.7in]{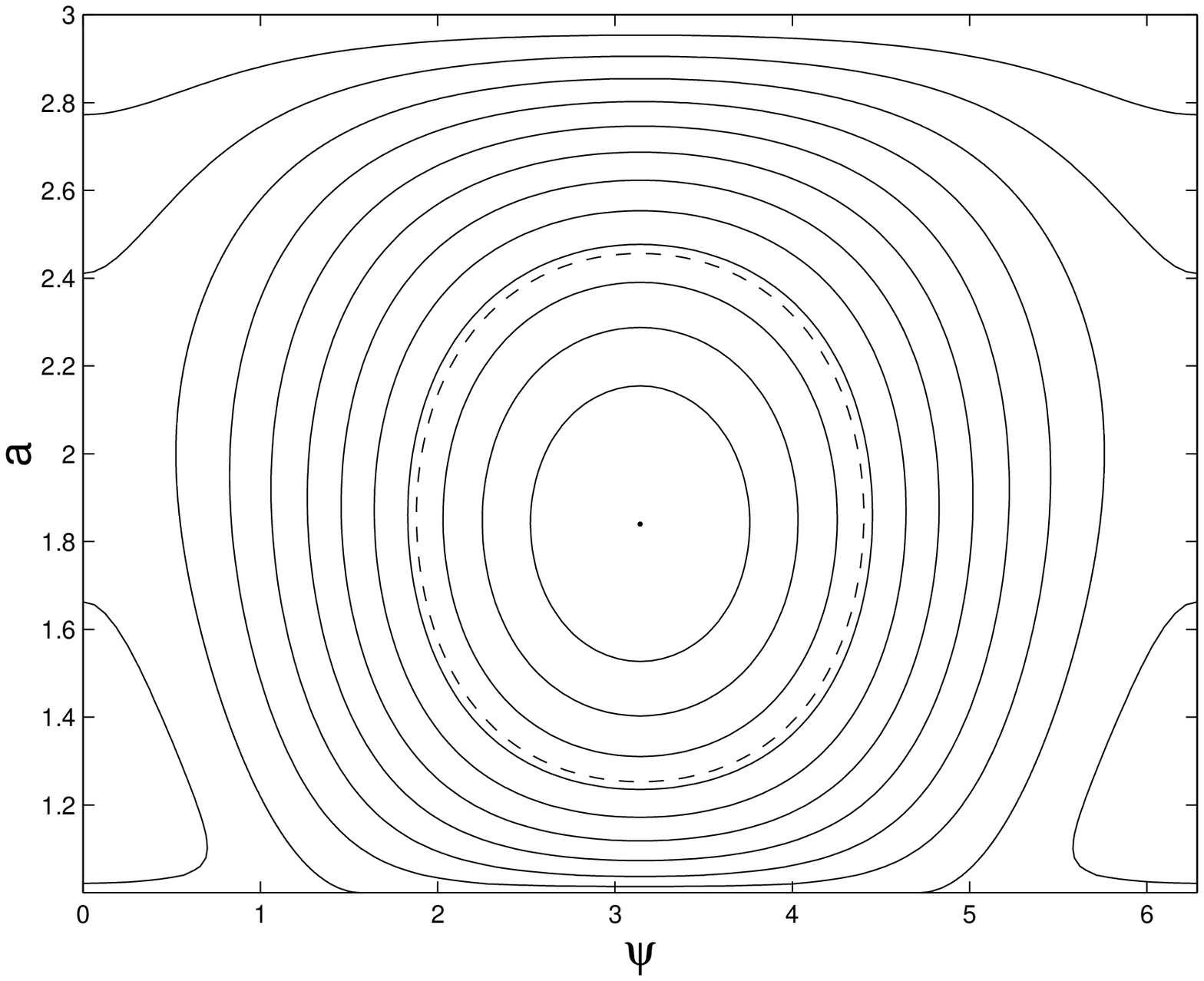}
\includegraphics[width=2.7in]{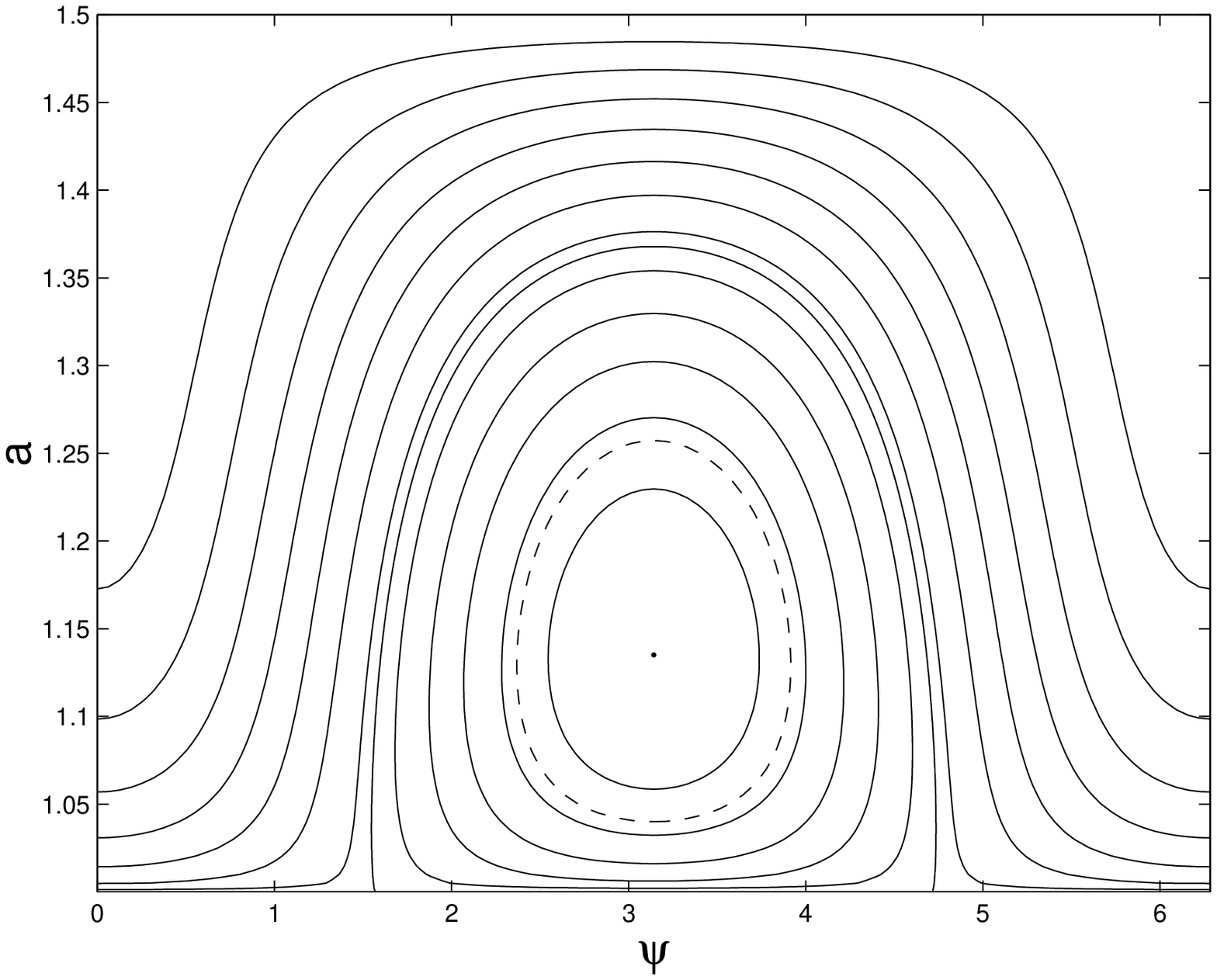}
\end{center}
\caption{Level sets of $H_0$ in the invariant plane $\cP_0$ for
(a): $c=5 \ (\eta=4)$; (b) $c=3 \ (\eta=2)$ and (c) $c=1.5 \ (\eta=0.5)$.
The dashed curve indicates the boundary of the orbits accessible 
from infinity, as described in the text.}
\label{fig:P_0}
\end{figure}

As noted above, the boundaries $a=c$ and $a=1$ of ${\cP_0}$ are
invariant, and the flow is singular on the latter, which contains two
degenerate saddle points at $\psi = \pi/2, \; 3\pi/2$. There is a
unique fixed point $(\pi,a^*)$ on $\psi = \pi$ surrounded by periodic
orbits which limit on heteroclinic orbit(s). As shown in the first
section of the Appendix, for all $c \; ( > 1)$, $(\pi,a^*)$ is a
saddle-center, with positive and negative real eigenvalues whose
eigenvectors point out of ${\cP_0}$. For $c < 2.214 \ldots$ this is
the only equilibrium; for $c > 2.214 \ldots$ two further fixed points,
a center-center and a saddle-center, appear on $\psi = 0$; restricted
to ${\cP_0}$ these are a center and a saddle, whose separatrices
interact with the stable and unstable manifolds of the degenerate
saddles on $a=1$ in a heteroclinic bifurcation~\cite{GH:83} at $c
\approx 3.21$ as $c$ continues to increase. $H_0$ takes its minimum
value $-c^3/3$ on $c=a$, its maximum at $(\pi,a^*)$, and the value
$-c^3/3 + c -1$ on $a=1$ and the invariant manifolds emanating from
it. Figure~\ref{fig:P_0} shows these distinct cases. On this
figure we also show the level set with Hamiltonian value equal to that
of a `pure' soliton stalled at infinity: $H(|Z| = \infty, V = 0, a =
1, \psi) = -c^3/3 + c(c - 1)$.  Since any incoming soliton with
nonzero speed has $H > -c^3/3 + c(c - 1)$ (see~\eqref{eq:H1}), this
curve bounds the set of {\em accessible\/} orbits on ${\cP_0}$: a disk
centered on $(\pi,a^*)$.

\subsection{Stable and unstable manifolds of ${\cP_{\infty}}$}
\label{sec:P_inf}

We first observe that we may define a local three-dimensional cross
section~\cite{GH:83} for the flow of~\eqref{eq:ode}: 
\beq
\Sigma_{\pi} =
\{ (Z,V,\psi,a) | a \in (a^*, c), \psi = \pi \}.
\label{eq:sigma_pi}
\eeq
 We verify that the
flow is transverse  to $\Sigma_{\pi}$ in the second section of the
Appendix.

Since the Hamiltonian~\eqref{eq:H1} is time-independent, its value is
conserved by solutions of~\eqref{eq:ode}, which are therefore
constrained to lie on three-dimensional surfaces
$$
H(V,\psi,p,I;c) = h = \text{const} ,
$$
determined by the initial data. The variables $p$ and $I$, defined below,  
denote
coordinates which are conjugate to coordinates $V$ and $\psi$. As
shown in the final section of the Appendix, this permits a further
reduction in dimension. Specifically, since the cross section
$\Sigma_{\pi}$ intersects level sets of $H$ transversely,
$\Sigma_{\pi}$ may be used as a two-dimensional cross section for the
flow  restricted to constant $H$ surfaces. It is on this cross
section that we will portray the stable and unstable manifolds
$W^{s,u}({\cP_{\infty}})$.

To approximate these manifolds we first add an artificial coupling
parameter $\mu$ to the Hamiltonian of Equation~\eqref{eq:H1}:
\beq
H = -\frac{c^3}{3} + (c-a) 
 \left( ca + V^2 - (c-a)  \sech^2 Z \right)
 - \mu 2 (c-a) \sqrt{a^2-1} \sech Z \cos \psi  .
\label{eq:Hmu}
\eeq
For the case at hand, $\mu = 1$, but we shall assume $\mu \ll 1$ and
perform a perturbative analysis, subsequently appealing to
continuation to extend to $\mu = 1$.  The variables $V$ and $\psi$ are
canonical `positions' for this Hamiltonian with conjugate `momenta'
$p=-2Z$ and $I = 2(a-\gamma) = 2(a-1)$, and $I$ and $\psi$ are
action-angle variables for the `second' degree of freedom. In these
canonical variables, the Hamiltonian~\eqref{eq:Hmu} assumes the form:
\beq
H = H_0(V,p;I) + \mu H_1(p;I,\psi) .
\label{eq:H0H1}
\eeq
The formal discussion of the Melnikov integral will refer to these
canonical variables, although in both the computations to follow,
and in numerical simulations, it is more convenient to
work with the original variables $(Z,V,\psi,a)$. 

For $\mu = 0$, the `unperturbed' Hamiltonian $H_0$ is independent 
of $\psi$, and the ODEs reduce to
\begin{subequations}
\begin{align}
\dot Z &= (c-a)V,\\
\dot V &= -(c-a)^2 \sech^2{Z} \tanh{Z} \\ 
\dot \psi &= \frac{c^2 - 2ca - V^2}{2} + (c-a) \sech^2{Z}  \label{eq:mu0_psidot}\\
\dot a &= 0 .
\end{align}
\label{eq:mu0_ode}
\end{subequations}
The position and velocity $Z$ and $V$ evolve as a particle in a
potential well, with the strength of the well dependent on $a=a_0$,
which is unchanging.  The solution set comprises bounded periodic
orbits, and unbounded orbits where $\abs{Z} \to \infty$ with finite
speed. In between is a separatrix, on which $Z=
\sinh^{-1}{\eta_0^{3/2} t}$, where $\eta_0=c-a_0$ is the amplitude of
the soliton.  In the unperturbed dynamics, the angular displacement,
determined by integration of~\eqref{eq:mu0_psidot} along the separatrix,
is given by
\beq 
\psi = \psi_0 + \frac{\eta_0^2 - a_0^2}{2} + \frac{1}{2\sqrt{\eta_0}}
\arctan{\eta_0^{3/2}} .
\label{eq:psi}
\eeq
The separatrices are homoclinic orbits to a periodic orbit $\beta(t)
\in {\cP_{\infty}}$ with $a = a_0$ and $(Z, V)= (\pm \infty, 0)$. Note
that unlike in~\cite{GHW:01} and related problems in celestial
mechanics (e.g.~\cite{DH:95}), in this case the generalized `position'
(soliton speed, $V$) goes to zero and the conjugate `momentum' (soliton
position, $Z$) goes to infinity.

We ask if any members of this continuum of homoclinic orbits persist
when $\mu \neq 0$. Now if $W^u (\beta)$
crosses the $p = 0$ axis at the point $(p=0,V,\psi=0,a,t=0)$, then, by the
symmetries~\eqref{eq:symm}, $W^s (\beta)$ must
also pass through this point. To establish existence of a (perturbed)
homoclinic orbit it therefore suffices to show that either $W^s (\beta)$ or
$W^u (\beta)$ intersects $p=0$. However, to demonstrate transverse
intersections requires a more delicate calculation, appealing to
perturbation theory for small $\mu$ and a result due to Melnikov, the 
derivation of which is outlined in the final section of the Appendix. 
Specifically, we have:

\begin{thm}
\label{thm:Mel}
Let $h > 0$ and $\Omega_0(t) = \pdiff{H_0}{I}$.  
Let $\{H_0,\frac{H_1}{\Omega_0}\}$denote the Poisson bracket\footnote{
$\{F,G\}\defeq\pdiff{F}{V}\pdiff{G}{p}-\pdiff{F}{p}\pdiff{G}{V}$}
of $H_0(V^0,p^0)$ and $H^1(V^0,p^0,\psi^0,I^0)/\Omega_{0}(V^0,p^0)$ evaluated
along $V^0(t)$ and $p^0(t)$.  Define the Melnikov function
\begin{equation} 
M(\psi_0) = \intinf \left \{H_0, \frac{H_1}{\Omega_0}\right\}
(V,p^0,\psi^0,I^0)(t) \ dt ,
\end{equation}
and assume that $M(\psi_0)$ has a simple zero and $\Omega_0(t) \neq 0$.  Then
for $\mu > 0$ sufficiently small, the Hamiltonian system has transverse
homoclinic orbits on the energy surface $H=h^0$.
\end{thm}
 
We note that the usual reduction process and Melnikov integral are
meaningful only as long as $\psi$ is monotonic with respect to $t$
($\dot{\psi} = \Omega \neq 0$), so that the global cross section
$\Sigma_{\psi_0}$ referred to in the Appendix may be defined.
However, Holmes~\cite{H:86} has extended this analysis to the case
where $\dot \psi$ is allowed to change sign in a bounded region in the
`middle' of the unperturbed homoclinic orbit, but under the condition
that $\psi$ be monotonic sufficiently close to the periodic orbit
$\beta$. Direct substitution of the unperturbed orbit $Z =
\sinh^{-1}{\eta_0^{3/2} t}$ into Equation~\eqref{eq:mu0_psidot} yields
\beq
\Omega_0 = \frac{1}{2} \left( \eta_0^2-a_0^2 + 
           \frac{\eta_0}{1+\eta_0^3t^2} \right) .
\eeq 
By our ansatz $\eta_0>0$, and so the condition that $\Omega_0$ not 
change sign throughout is
$$
(1 <) \  a_0 < \eta_0 \text{ or }  a_0 > \sqrt{\eta_0^2+\eta_0} ;
$$
however, if we appeal to~\cite{H:86}, we need only exclude a 
neighborhood of the degenerate set
$$
 a_0 = \eta_0
$$
to ensure that $\Omega_0 \neq 0$ near $\beta$.

We now sketch the computation needed to verify the remaining hypothesis of
Theorem~\ref{thm:Mel}. Since $H_1$ is $V$-independent, the Poisson
bracket reduces to
$$
\left\{H_0,\frac{H_1}{\Omega_0}\right\}_{(V,p)}
= \frac{1}{\Omega_0} \pdiff{H_0}{V}\pdiff{H_1}{p} - \frac{H_1}{\Omega_0^2}
\left(\pdiff{H_0}{V}\pdiff{\Omega_0}{p}-\pdiff{H_0}{p}\pdiff{\Omega_0}{V}
\right).
$$
After computing partial derivatives (cf.~\eqref{eq:mu0_ode}) and
substitution of the unperturbed separatrix solution in the Hamiltonian
of~\eqref{eq:Hmu}, some manipulations, and appeal to odd- and evenness
properties, the Melnikov integral may be written as:
\beq
\cM(\psi_0)= \eta_0^4\sqrt{a_0^2-1}  
\left(
\int_{-\infty}^\infty 
\frac{\left( -(\eta_0^2-a_0^2)g(t) +\eta_0 g^{-1}(t) \right)\sin{\Theta(t)}}
{\left(\left(\eta_0^2-a_0^2\right)g^2(t) +\eta_0\right)^2} \, dt \right)\sin{\psi_0} ,
\eeq
where
$$
g(t) = \sqrt{1+\eta_0^3 t^2} \text{ and }
\Theta(t) = \frac{\eta_0^2-a_0^2}{2}t + 
\frac{1}{2\sqrt{\eta_0}} \arctan{\eta_0^{\frac{3}{2}}t} .
$$
This integral cannot, in general, be computed explicitly, unless
$\sqrt{\eta_0}$ is rational. However, since the integrand is an
analytic function of $t$ with no essential singularities, it can
have only isolated zeros. Therefore, except for special values of
$(\eta_0,a_0)$, $\cM(\psi_0)$ has only simple zeros where  $\psi_0=
n\pi$, and Theorem~\ref{thm:Mel} implies that, for small values 
of $\mu$, there exist transverse homoclinic orbits to infinity.

\begin{figure}
\begin{center}
\includegraphics[width=3.3in]{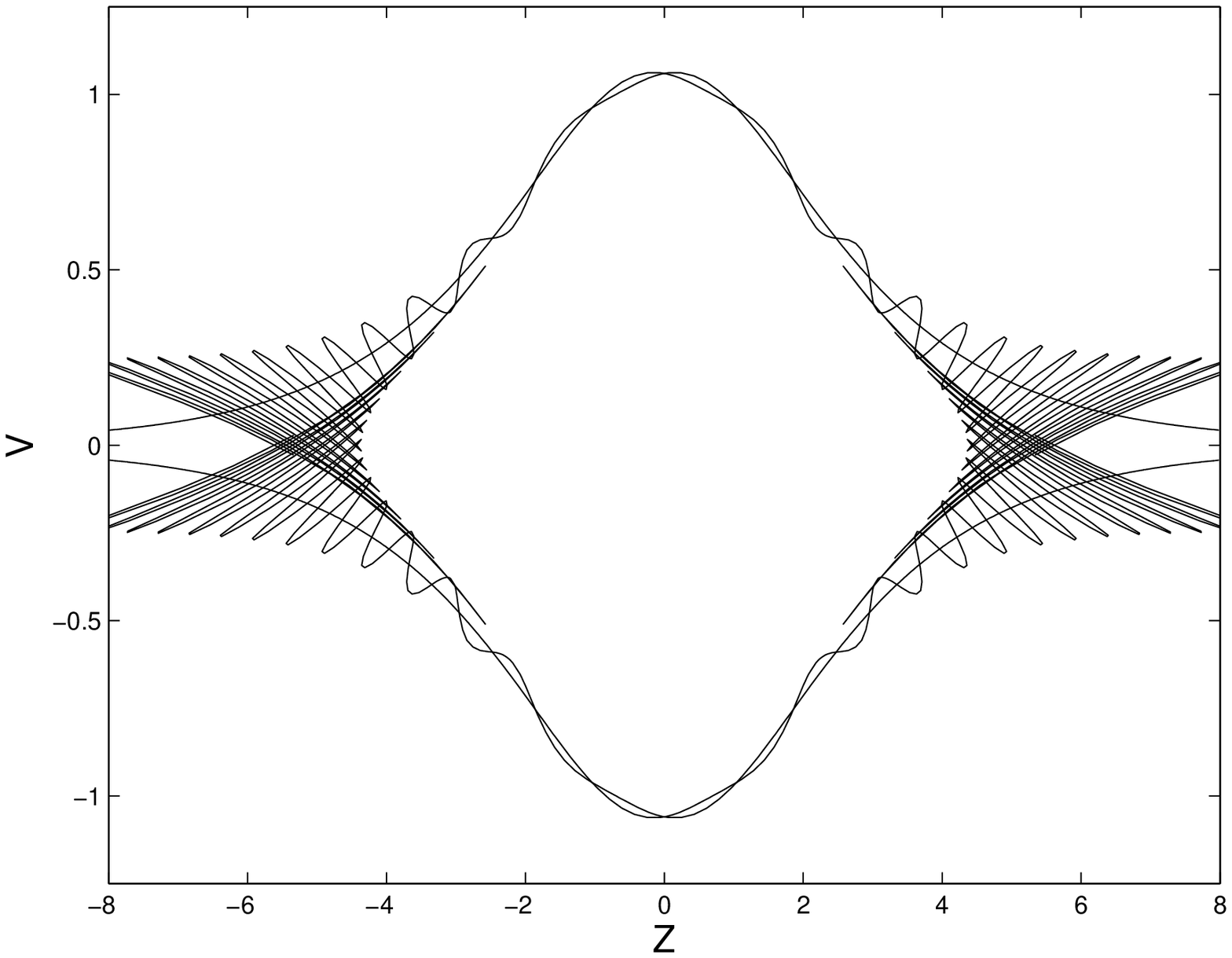}
\includegraphics[width=3.3in]{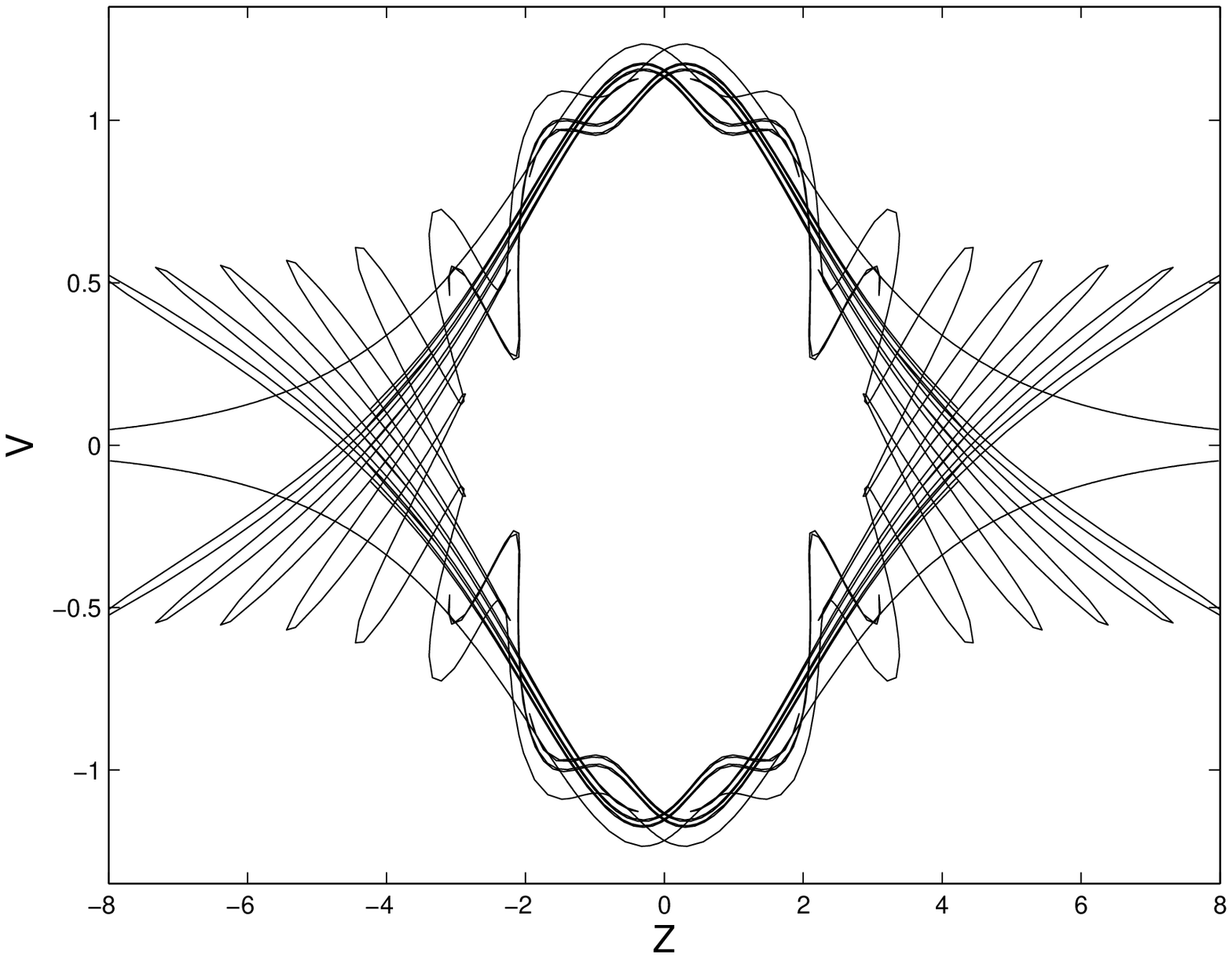}
\end{center}
\caption{ Stable and unstable manifolds of periodic orbits $\beta 
\in \cP_{\infty}$, shown via the Poincar\'{e} map defined on the 
cross section $\Sigma_{\pi}$. (a) $\mu = 0.25$; (b) $\mu = 0.5$, 
corresponding to the energy level for $\eta=4$ and $V=0.4$.  } 
\label{fig:Poincaremap}
\end{figure}

Figure~\ref{fig:Poincaremap} shows numerical computations of the
stable and unstable manifolds of periodic orbits in $\cP_{\infty}$ for
two values of $\mu$ in the same energy surface $H = -20.69$, corresponding
to the energy of the system with a soliton with $\eta=4$ and $V=0.4$ starting at $\abs{Z}=\infty$, with $a=0$.
They are illustrated as curves lying in the cross
section $\Sigma_{\pi}$ introduced at the beginning of this
section. (Since each periodic orbit $\beta$ in $\cP_{\infty}$ is a
one-dimensional circle, $W^s (\beta)$ and $W^u (\beta)$ are each
two-dimensional, and so intersect suitable cross sections to the flow
in one-dimensional curves.) The transverse intersections are
clear from the figure. As described in~\cite{GHW:01}, phase space transport
theory~\cite{RKW:90,W:92} may be used to analyse the capture,
transmission and reflection dynamics implied by transveral homoclinic
points such as those of Figure~\ref{fig:Poincaremap}. Here we provide
a brief review; for a more  complete explanation, see~\cite{GHW:01}. 

We consider $W^{u}(Z=-\infty)$, the unstable  manifold of $Z=-\infty$,
and $W^{s}(Z=\infty)$, the stable manifold of $Z=\infty$, which
intersect transversely in a point $q_{0}$ in the top center of the
figure.  (Both cases show such an intersection, although the
phenomenon is clearer in the lower figure.)  The union of the point
$q_{0}$,  the portion of  $W^{u}(Z=-\infty)$ to the left of $q_{0}$,
and the portion of $W^{s}(Z=\infty)$ to the right of $q_{0}$  form a
boundary between the upper and middle regions of the plane.  A similar
boundary exists in the lower half plane.  Stretching off to the left
from the point $q_{0}$ is a series of lobes lying in the upper region,
each of which is the image under the map $P_{\pi}$ of the lobe to its
left.  Between each such pair of lobes in the upper region, and below
$W^{u}(Z=-\infty)$, lies a lobe in the  middle region. Counting both
sets of lobes, the image of any given lobe under $P_{\pi}$ is the
second lobe to its right. In particular, the image of the nearest
upper region lobe to the left of $q_{0}$ is a lobe located in the
\textit{middle} region. Similarly, the image of the middle region lobe
immediately to the left of $q_{0}$ lies in the upper region. This
nearest upper lobe and the neighboring middle lobe form a
\textit{turnstile} through which phase space points are transported
from the upper to the middle region, and from the middle to the upper
region. A similar turnstile in the lower half plane
transports phase space between the lower and middle regions.

Trapping takes place when an initial condition that lies in the
sequence of lobes in the upper-left quadrant is mapped from the upper
region to the middle region. Reflection or transmission occurs because
eventually, as the interior lobes are successively stretched and
folded, the image of this point will, with probability one, lie in a
turnstile exit lobe (area preservation of the symplectic map $P_{\pi}$
guarantees that no open set contained in a preimage of an incoming
turnstile lobe can be trapped for all future time~\cite[Proposition
1]{GHW:01}). If the point's image exits into the upper region, it is
transmitted. If it goes into the lower region, it is reflected.  This
may be seen as an analogy with phase space transport in a two-bend
horseshoe map, as is shown in~\cite{GHW:01}. The iterated preimages of
the turnstile lobes forms a fractal structure that is responsible for
the multiple reflection and transmission windows of
Figure~\ref{fig:in_out_eta_4}.

\subsection{Stability of orbits on ${\cP_0}$}
\label{sec:P_0}

To determine the stable and unstable manifolds of ${\cP_0}$ we must
first determine the stability types of orbits within it, to
perturbations out of ${\cP_0}$.  By continuity, the periodic orbits
immediately surrounding the saddle center $(\pi,a^*)$ are also of
saddle type with respect to such perturbations, but the stability
types of other periodic orbits must be determined via Floquet
theory~\cite{Hart:64}.

On ${\cP_0}$ the ODEs reduce to:
\begin{subequations}
\begin{align}
\dot \psi &= \frac{c^2}{2} - ca + (c-a) +
\frac{(2a^2-ac-1)}{\sqrt{a^2 - 1}} \cos{\psi} , \\
\dot a &= -(c-a) \sqrt{a^2 - 1} \sin{\psi} .
\end{align}
\label{eq:P_odes}
\end{subequations}
Typical phase portraits of~\eqref{eq:P_odes} are shown in 
Figure~\ref{fig:P_0} above.

Consider the solution $\cS^*$ to the initial value problem of a
soliton starting from $\abs{Z}=\infty$ with finite velocity $V_\infty$
and zero energy in the defect mode, i.e.\ $a=1$ and
$c=\eta+1$\footnote{Practically, as in Section~\ref{sec:odesim}, the orbit is
started at some $|Z_0| \gg 1$ with $a = 1 + \varepsilon, \;
\varepsilon \ll 1$.}. $\cS^*$ is confined to the level set
\begin{equation}
H = -\frac{c^3}{3} + (c - 1) (c + V_{\infty}^2)
\stackrel{\rm{def}}{=} H_{\infty}
\label{eq:H_S*}
\end{equation}
of the conserved Hamiltonian and hence, if it approaches ${\cP_0}$,
can only interact with orbits having the same $H$-value. In
particular, since the maximum $H$ value for orbits on ${\cP_0}$ is
assumed by the fixed point $(\pi, a^*)$, there is a critical velocity
$V_\infty^{\rm max}$ above which the solutions $\cS^*$ have more
`energy' than any orbits contained in ${\cP_0}$, and thus must remain
bounded away from it. Similarly, the minimal value $H = -c^3/3
+ c (c - 1)$ of orbits $\cS^*$, assumed when $V_\infty = 0$, bounds
the set of accessible orbits on ${\cP_0}$, as  shown by the dashed curve on
Figure~\ref{fig:P_0}.

Consequently, for each $V_\infty \in [0, V_\infty^{\rm max})$, we find
a periodic orbit ${\cS_0} \in {\cP_0}$ with the same Hamiltonian value
$H = h_0$ as $\cS^*$ and determine its stability by examining the
linearization of the full system~\eqref{eq:ode} about ${\cS_0} =
(0,0,\psi_\cP(t),a_\cP(t))$. The stability of such an orbit is given
by the eigenvalues of the monodromy matrix: the fundamental solution
matrix of the linearized differential equation, evaluated at one period
of oscillation.  Let $\tilde{\cS_0} = (\tilde{Z}, \tilde{V},
\tilde{\psi}, \tilde{a})$ solve this linearized ODE, which is
block-diagonal, with the $(\tilde{Z}, \tilde{V})$ components
decoupling from the $(\tilde{\psi}, \tilde{a})$ components. The
eigenspace of the latter coincides with ${\cP_0}$ and hence they have
eigenvalues of unit modulus, as one expects from the integrable
structure of Figure~\ref{fig:P_0} (in fact $\lambda=1$ with
multiplicity two and a single eigenvector).

Perturbations perpendicular to ${\cP_0}$ satisfy
\beq
\begin{split}
 \diff{}{t} {\Tilde Z} &= (c-a_\cP(t)) \Tilde V , \\
 \diff{}{t} {\tilde V} &= -\left( (c-a_\cP(t))^2 +
 (c-a_\cP(t))\sqrt{a_\cP(t)^2-1}\cos{\psi_\cP(t)} \right) \tilde Z .
\end{split}
\label{eq:linP0}
\eeq
Since this may be written in the form
$$
\diff{}{t} \binom{\tilde{Z}}{\tilde{V}} = 
\begin{pmatrix}0 & A_{1,2}(t) \\ A_{2,1}(t) & 0 \end{pmatrix}
\binom{\tilde{Z}}{\tilde{V}} ,
$$
the Floquet theory for Hill's equation is
applicable~\cite{Hart:64}. The product of the Floquet eigenvalues must
be one, and their sum is given by the Floquet discriminant.  If this
is greater than two in absolute value, the periodic orbit is
hyperbolic, i.e. unstable; if less than two, the orbit is elliptic,
i.e. neutrally stable, transverse to ${\cP_0}$. We approximate these
discriminants by first numerically integrating the orbit
$(a_\cP(t),\psi_\cP(t))$ for one period, and then computing the
fundamental solution matrix for the linearized
system~\eqref{eq:linP0}, using interpolated data for the coefficients.

\begin{figure}
\begin{center}
\includegraphics[width=2.7in]{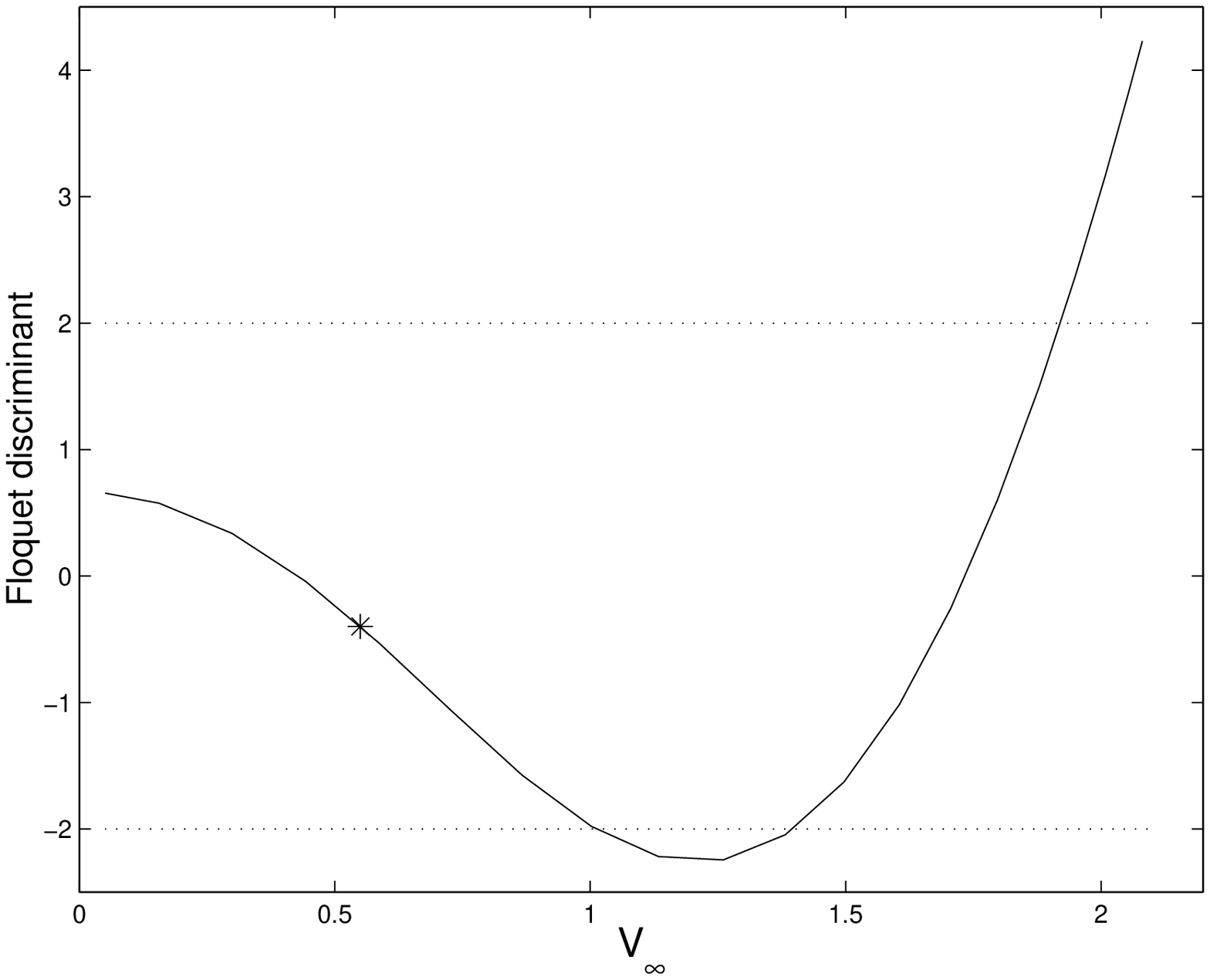}
\includegraphics[width=2.7in]{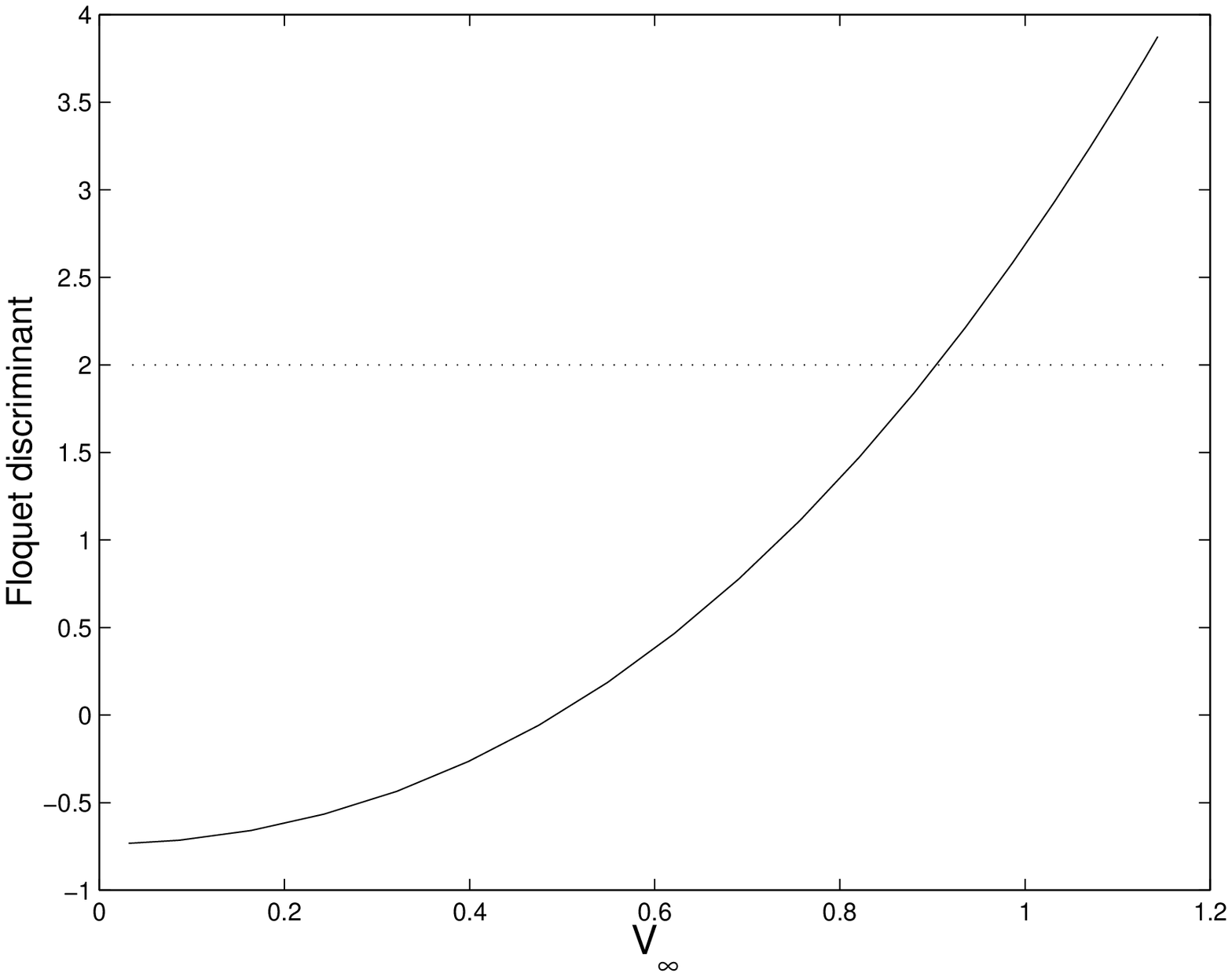}
\includegraphics[width=2.7in]{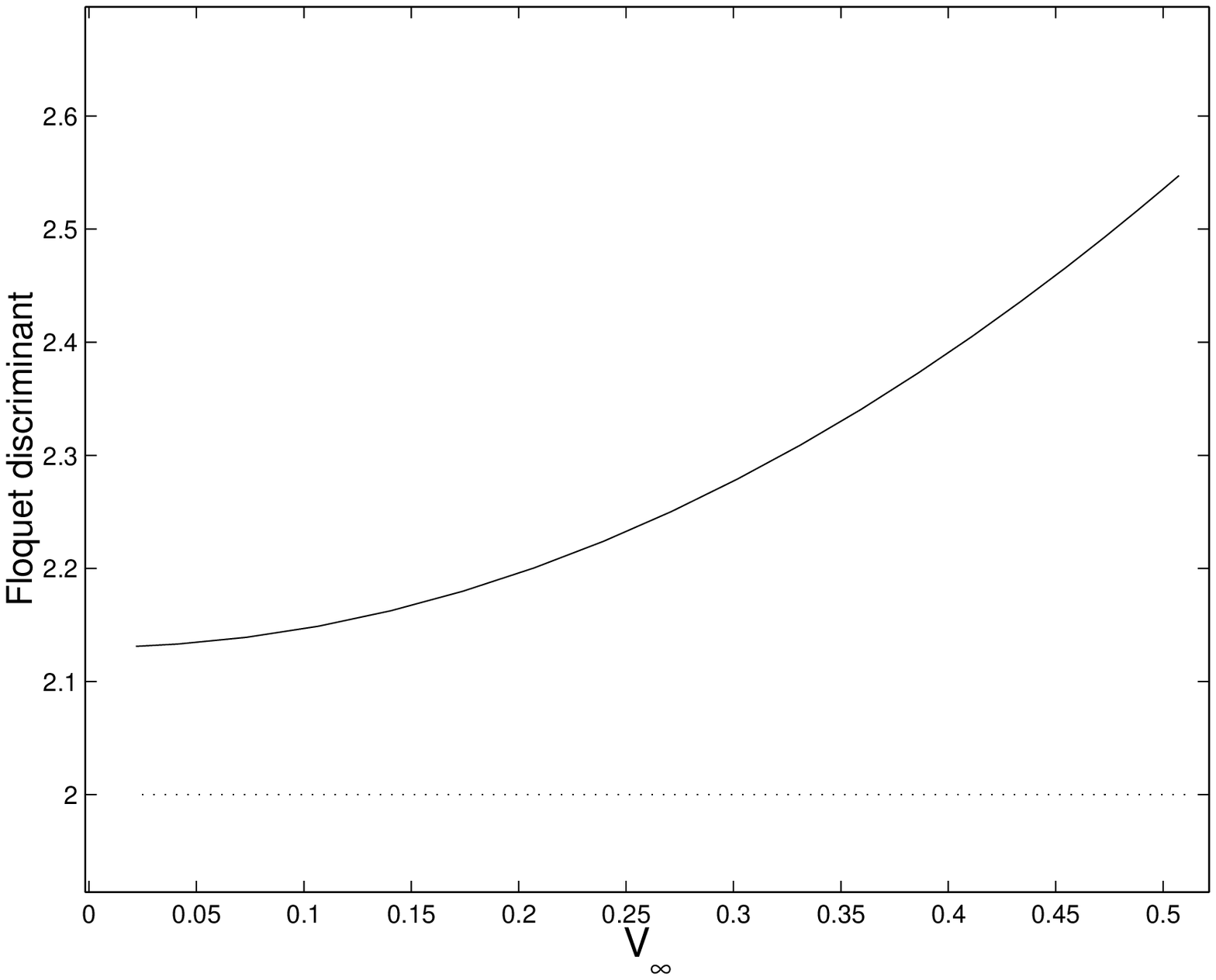}
\end{center}
\caption{The Floquet discriminant vs. incoming velocity for
$\eta = {4}, \, {2},\, \frac{1}{2}$ respectively (top to bottom).
In the top panel, the critical velocity separating capture 
from transmission is marked with an asterisk.} 
\label{fig:discrim}
\end{figure}

Since each periodic orbit in ${\cP_0}$ corresponds, via its Hamiltonian level,
to a velocity $V_\infty$ at $|Z| = \infty$, we plot in
Figure~\ref{fig:discrim} the Floquet discriminants as functions of $V_\infty$
for the three examples of Section~\ref{sec:smalleta}.  In the first case
($\eta = 4$) there are two regions each of stability and instability, and the
velocity range shown in Figure~\ref{fig:in_out_eta_4} corresponds to periodic
orbits of {\em elliptic\/} type, consistent with the intuition from
Figure~\ref{fig:captured_transmitted}, that solutions oscillate about
${\cP_0}$. The critical velocity dividing capture and transmission, identified
in Figure~\ref{fig:in_out_eta_4} is indicated by an asterisk. The second case
of $\eta = 2$ encompasses a range of stability and one of instability, and in
the third (and simplest) case, with $\eta = \frac{1}{2}$, all orbits are
hyperbolic. It is notable that $V_\infty^{\rm max} = 0.51$ in this case,
approximately equal to the critical velocity for transmission and reflection
seen in Figure~\ref{fig:in_out_eta_half} (but see the detailed analysis in
Section~\ref{sec:P_0mfd}, below).

Some interesting implications can immediately be drawn from the
stability types of orbits in ${\cP_0}$ implicit in
Figure~\ref{fig:discrim}. Recall that these periodic orbits correspond
to a state in which a soliton, stalled over the defect at $Z=0$,
exchanges energy periodically with the defect mode. Hence, if soliton
and defect parameters are chosen consistent with a stable region (eg.,
below $V_\infty \approx 0.9$ in case $\eta = 2$ and for $V_\infty < 1$ and
$V_\infty \in (1.4,1.8)$ in case $\eta = 4$), and the soliton is
initialized at the defect or somehow introduced into it, perhaps by
temporarily destabilizing the relevant orbit, it will remain trapped
under small perturbations. Such stable trapped states do not exist
for smaller $\eta$.

In the case $\eta=0.5$, the initial condition $\cS^*_c=(-\infty, V_c
=0.501, 1, \psi_0)$ lies on the same Hamiltonian surface as the fixed
point $(\pi,a^*)$. The stable and unstable manifolds of this fixed
point are only one dimensional and thus cannot separate reflected from
transmitted orbits.  However, the stable and unstable manifolds of
(the accessible disk on) ${\cP_0}$ are three dimensional, and are
consequently able to divide the phase space into disjoint regions. We
must therefore compute the stable and unstable manifolds of the
accessible periodic orbits on ${\cP_0}$ as well as of the
saddle-center $(\pi,a^*)$.

\subsection{Stable and unstable manifolds of ${\cP_0}$}
\label{sec:P_0mfd}

Appealing to the symmetries of~\eqref{eq:symm} and the fact that the
orbits of interest are reflection-symmetric about $\psi = \pi$, we
need only compute one of the four branches of $W^{s,u}({\cS_0})$ for
each of the saddle type periodic orbits ${\cS_0} \in {\cP_0}$. To do
this we first compute each periodic orbit $\cS_0$ starting at a point
$(a,\psi)=(a_0,\pi)$ where $a_0 > a^*$, the saddle-center. We
interpolate these with 64 equally-spaced points (with respect to
time). At each of these 64 points $(a_0,\psi_0)$, the fundamental
solution matrix is computed as in Subsection~\ref{sec:P_0}. Fourier
interpolation is used to compute the coefficients $(a_\cP,\psi_\cP)$
at intermediate times, so the orbit $\cS_0$ need only be computed
once. At each point on the periodic orbit, we compute the unstable
eigenvector of the monodromy matrix, $\vec v_0 = (Z_0,V_0)^T$.  We
normalize it so that $| \vec v_0 | = 10^{-5}$ and solve the full
system~\eqref{eq:ode} of ODE's with initial conditions
$(Z_0,V_0,a_0,\psi_0)$, stopping when $\abs{Z} =20$.

Let $W^s({\cP_0}) = \cup_{h_0} W^s({\cS_0})$ denote the set of stable
manifolds of the accessible hyperbolic orbits in
${\cP_0}$. $W^s({\cP_0})$ is three dimensional, and is locally (near
$|Z| = 0$) foliated by two-dimensional cylinders, each of which is a
local stable manifold of some ${\cS_0}$. We therefore expect
$W^s({\cP_0})$ to intersect the three-dimensional cross sections of
initial data $\Sigma_{\pm} = \{(V,\psi,a) | Z = \pm Z_0, |Z_0| \gg 1
\}$ in two-dimensional sets, which should in turn separate sets of
initial data giving rise to solutions that pass the defect from those
reflected by it.

\begin{figure}
\begin{center}
\includegraphics[width=4in]{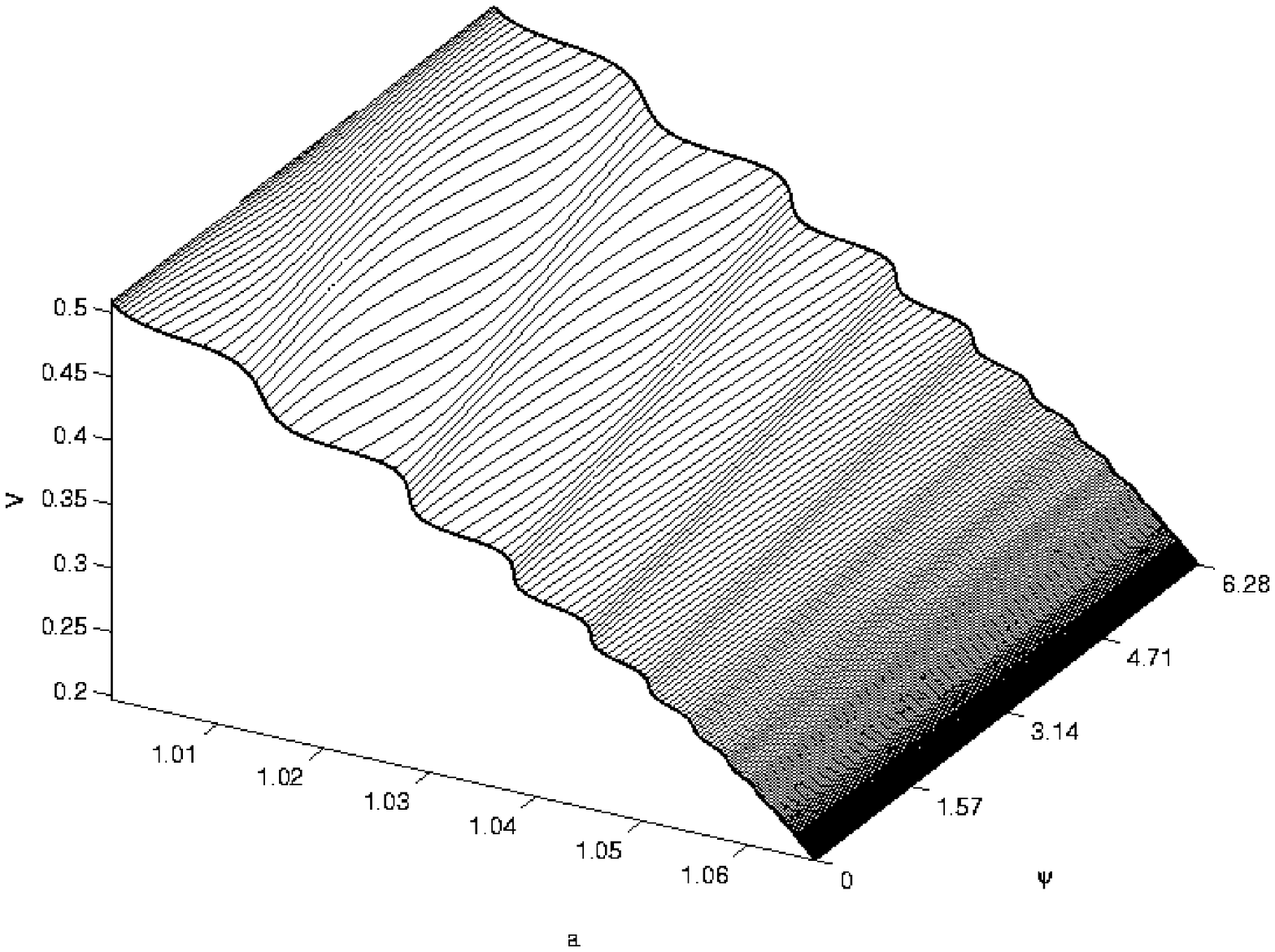}
\includegraphics[width=4in]{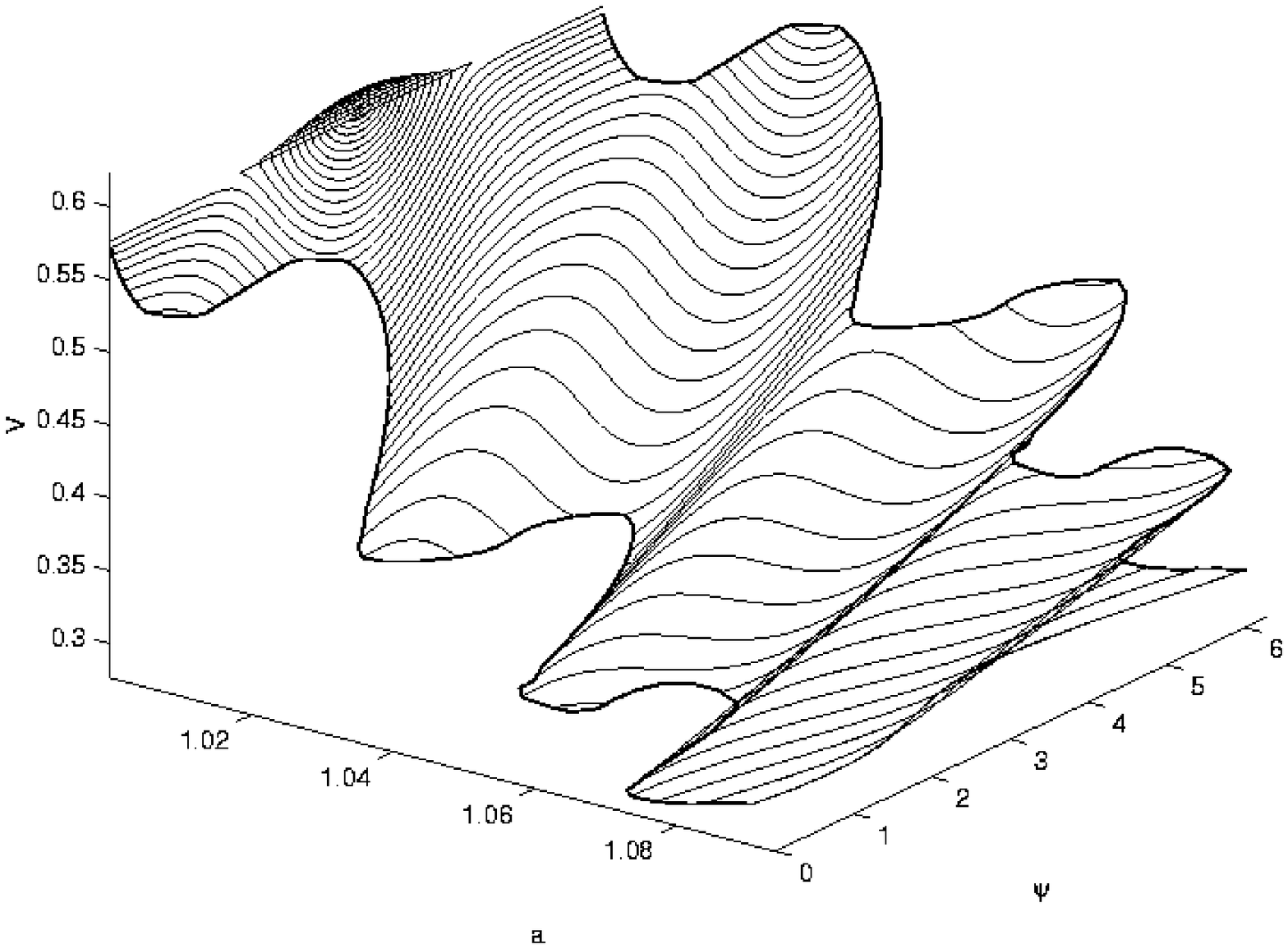}
\end{center}
\caption{Part of the intersection of the stable manifold of
${\cP_0}$ with $\Sigma_{-}$. (Top) $\eta = 0.5 \; (c = 1.5)$;
(bottom)  $\eta = 0.75 \; (c = 1.75)$.}
\label{fig:small_eta_mfd}
\end{figure}

Figure~\ref{fig:small_eta_mfd} shows the results of computations for
the third of the three cases of Section~\ref{sec:odesim}: $\eta =
0.5$, and for a further case, with slightly larger $\eta = 0.75$. Note
that the sets $\Sigma_{\pm} \sim \RR \times S^1 \times {\cI}$ are
periodic in $\psi$. As might be expected from the experiment of
Section~\ref{sec:smalleta}, near $a=1$ the surface $W^s({\cS_0}) \cap
\Sigma_{-}$ is a graph over the $(\psi,a)$ annulus: all orbits
starting at points below it are reflected, and points starting above
it are transmitted. Further from $a=1$ the surface develops folds;
these become more pronounced for higher $\eta \; (c)$, as in the lower
panel of Figure~\ref{fig:small_eta_mfd}. The surface describes the
critical velocity as a function of phase $\psi$ and amplitude $a-1$ of
the defect mode. Note the weak phase dependence, particularly as $a
\rightarrow 1$, and that the surface approaches $a=1$ at $V \approx
0.51$ in the case $\eta = 0.5$ (upper panel); this is the critical
velocity found in Section~\ref{sec:smalleta}. Initial data on this
surface corresponds to trapping (recall that the accessible orbits in
$\cP_0$ correspond to solitons pinned at the defect).

In interpreting this figure it is helpful to note the following
facts. Individual two-dimensional components $W^s({\cS_0})$ of
$W^s({\cP_0})$ intersect sections at $Z = \pm Z_0$ for small $|Z_0|$
in homotopically trivial (contractible) circles, but as $|Z|$ (and the
time of flight) increases, particular solutions belonging to
$W^s({\cS_0})$ can pass arbitrarily close to the degenerate saddles on
$a=1$ at $\psi = \pi/2, 3\pi/2$ (cf. Figure~\ref{fig:P_0}
(bottom)). This effectively separates neighboring solutions and
stretches their phase ($\psi$) angles over a range exceeding $2
\pi$. The result is that the corresponding sets $W^s({\cS_0}) \cap
\Sigma_{-}$ extend around the $S^1$ component in a homotopically
nontrivial manner. Only those components of $W^s({\cP_0})$ very close
to $W^s(\pi, a^*)$ remain contractible; these can be seen in
Figure~\ref{fig:small_eta_mfd} near $\psi=\pi, a = 1$.

The `outer' (lower $V$, higher $a$) boundary of the computed portion
of  $W^s({\cS_0}) \cap \Sigma_{-}$ is limited by numerical issues: it
is impossible to compute with uniform accuracy as velocities approach
zero, since the time of flight grows without bound; however, it
appears that velocities do decrease to zero as $a$ increases. For
example, numerical experiments like those of Section~\ref{sec:smalleta}
indicate that all orbits launched with positive velocities, no matter
how small, and $a > 1.1$,  are transmitted. The surface therefore
intersects $V=0$.

We have been unable to reliably compute invariant manifolds of $\cP_0$
for the medium and large $\eta$ cases. Preliminary studies suggest
that, as $c$ increases, the set $W^s({\cP_0}) \cap \Sigma_{-}$
`separates' from the plane $a=1$, so that nearby initial data all lie
in the transmission zone (cf. Figure~\ref{fig:in_out_eta_2}, which
indicates that all solitons are transmitted for $\eta = 2$, regardless
of their initial velocities). However, as $c$ continues to increase,
the stable manifold $W^s({\cP_{\infty}})$ evidently invades
$\Sigma_{-}$, leading to the complex behavior of
Figure~\ref{fig:in_out_eta_4}. In
particular, the increased folding of $W^s({\cP_0}) \cap \Sigma_{-}$ as
$\eta$ (or $c$) increases evident in Figure~\ref{fig:small_eta_mfd} is
consistent with the existence of a fine (fractal) structure suggested
by Figure~\ref{fig:in_out_eta_4}. Since $W^s({\cP_{\infty}})$
cannot intersect $W^s({\cP_0})$, we conjecture that, as
$W^s({\cP_{\infty}})$ invades $\Sigma_{-}$, it must `align' with the
latter (folded) manifold, producing (infinitely) many regions of
transmission and reflection on any vertical line in $\Sigma_{-}$ above
the $(a, \psi)$-plane.

\section{Interpretation and Summary}
\label{sec:summary}

In this paper, we have derived a finite
dimensional model for soliton-defect mode interactions in a nonlinear
Schr\"odinger equation with a point defect. Following~\cite{FPM:94},
and allowing for a fully nonlinear defect mode, which by itself is an
exact solution, we derive a three-degree-of-freedom Hamiltonian system
that describes the evolution of amplitudes and phases of the soliton
and defect mode, and the position and velocity of the former. Allowing
a nonlinear defect mode is important, since it permits resonant energy
transfer to occur over a range of soliton amplitudes.
However, only these two modes are represented; in particular, radiation
to the continuum is ignored, and multiple solitons are disallowed.

The resulting ODEs may be further reduced to two degrees of freedom,
since in addition to the Hamiltonian, a second quantity, corresponding
to the total energy in the two modes, is also conserved. While this
system is rather complex, and indeed is nonintegrable in certain
parameter ranges (cf. Section~\ref{sec:P_inf}), it possesses
two-dimensional invariant subspaces filled with periodic orbits, whose
stable and unstable manifolds partially determine the global structure
of solutions. We use this system to investigate the reflection,
transmission, and trapping of solitons launched `from infinity,' by
the defect, concentrating on the case in which the energy initially
all resides in the soliton. Numerical simulations of the model ODE's
reveal three basic types of behavior:

\begin{enumerate}

\item For large initial soliton intensities, there is a critical velocity
above which all solitons are transmitted; below this, a complex
structure of reflection and transmission bands exists, separated by
trapping regions that apparently are of measure zero.  The capture and
eventual transmission or reflection is explained by phase-space
transport via turnstile lobes formed from parts of the stable and unstable
manifolds of orbits `at infinity,' corresponding to uncoupled oscillations
of the defect mode and a distant soliton.

\item For moderate initial soliton intensities, all solitons are split
into a transmitted part and a captured defect mode.  The transmitted
part travels to the right monotonically, giving up a fraction of its
energy to the defect mode.  The amount of energy transferred to the
defect mode is a decreasing function of  incident velocity. 

\item For small initial soliton intensities, reflection or transmission
occurs for almost all initial velocities. Specifically, a unique
critical velocity exists for each initial phase and defect amplitude
below a certain level; this represents initial data on the stable
manifold of a subset of periodic orbits, each of which corresponds to
the soliton stalled over the defect, periodically exchanging energy
with the defect mode.  This can be explained by the stable and 
unstable manifold of the manifold $\cP_{0}$, which divide the
phase space into two regions.

\end{enumerate}

We now compare this to the behavior of the
numerically computed solutions to the original PDE:
 
\begin{enumerate}

\item For large initial soliton intensities, there also exists a
critical velocity separating solitons which are captured from those
which pass by the defect with little interaction. The ODE prediction
of $v_c = \eta V_c = 2.21$ (with $\eta = 4$) is in error by some $24
\%$ compared to the PDE simulations, $v_c = 1.78$, and unlike
solutions to the ODE, captured PDE solitons do not eventually escape.
The reduced ODE system is Hamiltonian, and thus, by a variant of the
Poincar\'e recurrence theorem, any solution for which $Z$ is unbounded
as $t\to-\infty$ must also have unbounded $Z$ at a later time, with
probability 1. In the full PDE, however, radiation damping plays a
role; there are radiative modes which carry energy away from the
defect mode. These radiation modes are not incorporated in our Ansatz;
as in other problems (cf.~\cite{GHW:01,SW:99}), their inclusion is
expected to yield a collective coordinate ODE reduction  with
corrections which can play the role of damping. This damping
corresponds to energy transfer from the soliton-defect mode subsystem
to the  radiative `heat bath.' While the finite dimensional
Hamiltonian  ODE reduction leads to trapping for a set of data of
measure zero, for the reduction perturbed by damping, a set of
data of positive measure is trapped.

The ODE model displays two behaviors for solitons below the critical
velocity.  Figure~\ref{fig:in_out_eta_4} reveals the existence
of reflection windows: initial velocities for which the soliton
returns to minus infinity with the same intensity and the opposite
velocity it started with.  Between these resonance windows are chaotic
regions, where the soliton may oscillate near the defect any number of
times before being ejected, with apparently random velocity.
Reflection windows are a common feature of ODE models of this
type~\cite{FKV:92,AOM:91} as well as in some PDEs describing
soliton--defect and soliton--soliton interactions~\cite{CSW:83,YT:01}.  
However, the fractal structure of the reflection/transmission
windows is seen only in the ODE models and not in the PDEs.  
Radiation damping becomes important when the soliton stays in
the neighborhood, and eliminates the chaotic behavior.
It is shown in~\cite{GHW:01} that the chaotic behavior can be
eliminated by the inclusion of appropriate radiation terms in the ODE
ansatz, which leads to damping in the ODE.

In numerical simulations of NLS soliton--defect interactions, no
reflection windows have ever been found via PDE simulations.  It
appears from our simulations, and from a form of post-processing of
the simulations to be described momentarily, that energy may be
transferred from the soliton to the defect mode, but not vice-versa,
so that once a defect mode is created, it never gives up its energy to
the soliton.  In the numerical post-processing, we calculate the six
ODE parameters of~\eqref{eq:Lagrangian} by minimizing the distance
between the numerical PDE solution and the ODE ansatz defined
by~\eqref{eq:u_S} and~\eqref{eq:u_B}.  In this analysis, for captured
solutions $\eta$ decreases to zero as $a$ grows, so that the soliton
is destroyed as the defect mode is created.  This is in contrast with
the sine-Gordon simulations of~\cite{FKV:92} in which reflection
windows are seen.  The difference lies in the fact that the two
interacting modes in the sine-Gordon experiments are `topologically'
distinct.  In that case the soliton is a kink, which approaches two
different limits as $x\to \pm \infty$, while the defect mode is
exponentially localized.  The kink is defined solely by position and
velocity, and has no amplitude parameter equivalent to $\eta$.  When
the kink transfers energy to the defect mode, it is not destroyed,
since it still must satisfy the boundary conditions at  $\pm
\infty$. This topological constraint forces it to persist, so that
energy stored in the defect mode may be (re-)converted  to kinetic
energy that pushes the kink away. In the NLS, in contrast, there is no
such topological barrier, since the soliton decays to zero at both
extremities. It can therefore transfer all its energy to the  defect
mode and cease to exist. No soliton-like structure need persist  to
absorb energy from the defect mode.

This points to a criterion we have not seen mentioned when evaluating
the effectiveness of collective coordinate ansatzen. If two modes
included in an ansatz can become `far from orthogonal'  in some
parameter regime, in that they may be highly correlated or may be used
to represent the same information, then collective coordinate methods
may give misleading results for solutions that approach these
parameter regimes.

\item For intermediate intensities, and specifically $\eta=2$, the
long time behavior consists of a captured defect mode and a
transmitted soliton.  The faster the soliton's initial approach, the
less energy is captured by the defect mode.  Comparing
Figure~\ref{fig:captured_vs_v_in} with the lower graph  of
Figure~\ref{fig:in_out_eta_2}, we see this `capture efficiency'
phenomenon in both cases. In neither case does a critical velocity
separate captured from transmitted solitons.

\item For small intensities, the soliton is split into three parts: reflected,
transmitted, and captured components.  A prediction is made in~\cite{CM:95}
about the amount of energy in each of these three modes, although no
comparison is made in numerics. Since the ansatz~(\ref{eq:u_S}-\ref{eq:u_B})
permits only two modes, it cannot possibly capture all of this behavior.  A
three-mode ansatz, including two solitons and a defect mode, was constructed,
but not found to be useful in studying this system.  Nonetheless at very
small (resp. large) initial velocities, the soliton is almost entirely
reflected (resp. transmitted), so the two-mode ansatz is reasonable.  In
these cases, the initial condition lies squarely to one side or the other of
the stable manifold illustrated in Figure~\ref{fig:small_eta_mfd}, and the
ODE and PDE simulations compare reasonably well.  At
intermediate velocities, the solution reaches a state shown in
Figure~\ref{fig:small_eta_recon}, which is a reconstruction of the ansatz
solution from the ODE parameters.  In this figure, the solution appears as a
soliton cleaved in two by a defect mode.  In the full PDE, the two halves of
the soliton component would separate and proceed in opposite directions.
In the ODE reduction, they are unable to do that: the ansatz ~(\ref{eq:u_S}-\ref{eq:u_B}) 
constrains recombination into a single soliton.

\end{enumerate}

\begin{figure}
\begin{center}
\includegraphics[width=4in]{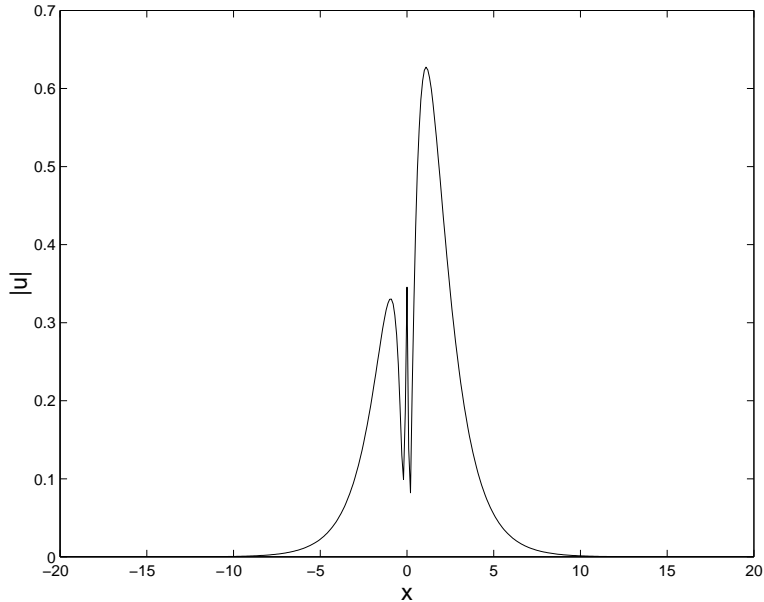}
\end{center}
\caption{Reconstruction of the spatial structure of solutions 
from the two-mode ansatz~(\ref{eq:u_S}-\ref{eq:u_B}), compared with 
full PDE solutions, for orbits reflected and transmitted in the 
case of $\eta = 0.5$. }
\label{fig:small_eta_recon}
\end{figure}

These comments illustrate what we believe to be rather general issues
relevant to understanding the successes and failures of
finite-dimensional or collective coordinate ODE reductions in
reproducing PDE dynamics.  Of course any correspondence between the
solutions to a PDE and its variational ODE approximation depends on
the assumption that the PDE solution remains close to the ansatz used
in the approximation.  This may or may not happen, and it is risky to
draw quantitative information from the ODE model, for example
regarding how solutions of the PDE depend on a certain parameter that
is varied.  Indeed, many such  studies amount to numerically solving
both the ODE and PDE, and noticing that the behavior is similar.

A more nuanced approach that is advocated in this paper and others, is
that  the ODE can be used to illuminate a mechanism that underlies a
behavior seen in numerical simulations of the PDE.  For example, we
have seen that the existence of a critical velocity in the PDE can be
explained by finding  separatrices in the reduced ODE dynamics.

We may even take this a step further.  An eventual goal of this
research is to understand the behavior of gap solitons interacting
with defects in Bragg gratings~\cite{GSW}.  The derivation of a
variational ODE for that system is complicated by the fact that gap
solitons possess internal modes, which would require additional
degrees of freedom in any collective coordinate ansatz, as
in~\cite{FKV:92a}.  Nonetheless, NLCME gap solitons interacting with
defects share many qualitative behaviors with the ODEs~\eqref{eq:ode}
derived in this study.  We may draw a bifurcation diagram much like
Figure~\ref{fig:L_v_freq} for this PDE.  Further, low amplitude gap
solitons are either coherently transmitted or reflected, whereas high
amplitude gap solitons are captured when sufficiently slow, and pass
by the defect if they have enough kinetic energy.  Both of these
behaviors were seen in the numerical experiments of
Section~\ref{sec:odesim}.  We may postulate that mechanisms similar to
those seen in section~\ref{sec:ode_analysis} are responsible for the
``resonance'' between solitons and defects described in~\cite{GSW}.

\vspace{0.5cm}

\noindent {\bf Acknowledgements:} \ RG was supported by NSF
DMS-9901897 and Bell Laboratories/Lucent Technologies under the
Postdoctoral Fellowship with Industry Program and by NSF DMS-0204881.
PH was partially supported by DoE: DE-FG02-95ER25238. We thank
D. Pelinovsky for stimulating discussions early in this project. Parts
of this paper were presented at `Mathematics as a Guide to the
Understanding of Applied Nonlinear Problems,' a conference in honor of
Klaus Kirchg\"{a}ssner's 70th birthday, Kloster Irsee, Germany, Jan
6-10, 2002, and an early version appeared in the informal festschrift
for that meeting, edited by H-J. Kielh\"{o}fer, A. Mielke, and
J. Scheurle.

\section*{Appendix: Detailed calculations}
\label{Apdx}

\subsection*{Fixed points on ${\cP_0}$}
\label{Apdx-fp}

We verify the claim that the fixed points $(\psi,a) = (\pi,a^*)$ on
${\cP_0}$ are saddle centers. The center behavior within ${\cP_0}$
follows from the structure of the restricted Hamiltonian~\eqref{eq:HP0}, 
and behavior transverse to ${\cP_0}$ is determined by
the linearization given in~\eqref{eq:linP0}, evaluated at $(\pi,a^*)$.
The resulting (constant) matrix has zero trace and determinant
\beq
 D = (c - a^*)^2 \left( (c - a^*) - \sqrt{(a^*)^2 -1} \right) 
\stackrel{\rm{def}}{=} (c - a^*)^2 \tilde{D}(a^*) ,
\label{eq:det}
\eeq
so, provided $\tilde{D}(a^*) < 0$, the remaining eigenvalues $\lambda =
\pm \sqrt{-D}$ are real, implying hyperbolic saddle type behavior in
the $(Z,V)$ directions. Note that $\tilde{D}(a^*) < 0$ iff
\beq
a^* > a_c = \frac{c^2 + 1}{2c} .
\label{eq:bnd}
\eeq
From~\eqref{eq:P_odes} the fixed point value $a^*$ is given by 
solution of
\beq
L(a) \stackrel{\rm{def}}{=} \frac{c^2}{2} + c - (c + 1) a =
 \frac{(2a^2 - ca - 1)}{\sqrt{a^2 - 1}} \stackrel{\rm{def}}{=} R(a) .
\label{eq:root}
\eeq
Now $L(a)$ is monotonically decreasing and $R(a)$ is monotonically 
increasing (in the range $a \in (1,c)$), so if we can show that
$L(a_c) > R(a_c)$, it follows that $a^* > a_c$ and hence that
$\tilde{D}(a^*) <0 $, as required. But
\[
L(a_c) = \frac{(c^2 - c - 1)}{2c} \ \ {\rm{and}} \ \
R(a_c) = - \frac{1}{c} ,
\]
and so the claim is true.

\subsection*{The cross section $\Sigma_{\pi}$}
\label{Apdx-xsec}

To verify that $\Sigma_{\pi} = \{ (Z,V,\psi,a) | a \in (a^*, c), 
\psi = \pi \}$ is a cross section for the flow it suffices to show that
$\dot{\psi} \neq 0$ on $\Sigma_{\pi}$. From~\eqref{eq:psidot} we have
\begin{eqnarray}
\dot{\psi}|_{\psi=\pi} &=& \frac{c^2-2ca}{2} - \frac{V^2}{2} + 
(c-a) \sech^2{Z} - \frac{(2a^2 - ca - 1)}{\sqrt{a^2 - 1}} \sech{Z} 
  \nonumber \\
&=& - \frac{V^2}{2} + (1 - \sech Z) \left[ \frac{c^2-2ca}{2}
  - (c-a) \sech Z \right]   \nonumber \\
& & \hspace{0.5cm} - \left( \frac{(2a^2 - ca - 1)}{\sqrt{a^2 - 1}} 
- \left[ \frac{c^2-2ca}{2} + (c-a) \right] \right) \sech Z   \nonumber \\
&\le& (1 - \sech Z) \left[ \frac{c^2-2ca}{2} - (c-a) \sech^2{Z} \right]
\nonumber \\
&& \hspace{0.5cm} - [R(a) - L(a)] \sech Z ,
\label{eq:xsec1}
\end{eqnarray}
where $R(a)$ and $L(a)$ are defined in~\eqref{eq:root}. Now $\sech Z < 1$ 
for $|Z| < \infty$, so the sign of the leading term in~\eqref{eq:xsec1} 
is determined by the expression in square brackets. The second term of 
this is always negative for $1 < a < c$ and the first is also negative 
for $a > a^*$, since $a^2 > a_c > c/2$, as shown above.  Finally, since
$R(a)$ is monotonically increasing and $L(a)$ monotonically 
decreasing and $a > a^*$, the point at which $R(a^*) = L(a^*)$, the 
last term is also strictly negative. We conclude that
$\dot{\psi} <  0$ on $\Sigma_{\pi}$, as required.

\subsection*{Reduction and the Melnikov function}
\label{Apdx-Mkov}

We summmarise the modified reduction procedure and Melnikov
calculation developed by Holmes and Marsden~\cite{HM:84} for two
degree-of-freedom Hamiltonian systems in the form~\eqref{eq:H0H1}, in
which the frequency of the action-angle mode depends upon the phase
variables in the other degree of freedom. (The procedure is also
outlined in~\cite{MH:88}, where it is applied to Kirchhoff's equations
for equilibria of an elastic rod.) As in Melnikov's `standard'
method~\cite{M:63,GH:83}, transverse intersections of stable
and unstable manifolds of a perturbed system are found by examining the
zeros of an integral computed along the homoclinic orbit of the
unperturbed system.

Consider the perturbed two-degree-of-freedom system with Hamiltonian 
\beq
H = H_0(V,p;I) + \mu H_1(V,p;I,\psi) = h \; (= \rm{const.}) ,
\label{eq:H0H1app}
\eeq
and let
\beq
\Omega = \pdiff{H_\mu}{I} = \pdiff{H_0}{I} + O(\mu) = \Omega_0 + O(\mu) .
\label{eq:omega}
\eeq
Then, provided $\Omega > 0$, Equation~\eqref{eq:H0H1app} may be
inverted and solved for $I$ in terms of $V$, $p$, $\psi$ and the
constant $h$. As shown in~\cite{HM:84}, we may therefore eliminate $I$
and replace time by the conjugate variable $\psi$ and write the
reduced three-dimensional system on the constant energy surface as a
periodically forced single degree of freedom system with Hamiltonian
$-I(V,p,\psi;L,h)$, and evolution equations
\beq 
V'= - \pdiff{I}{p}, \quad p' =\pdiff{I}{V},
\label{eq:reduced}
\eeq
where $(\cdot)'$ denotes $d/d\psi (\cdot)$. This implies conservation
of three-dimensional phase-space volume on the constant energy
manifolds, and area preservation in the two-dimensional Poincar\'e
maps defined below.  Moreover, in~\cite{HM:84} it is shown that the
reduced Hamiltonian  $I$ of~\eqref{eq:reduced} may be written
\beq
I = \cI_0(V,p;L,h) + \mu \cI_1(V,p,\psi;L,h) + O(\mu^2),
\label{eq:Iseries}
\eeq
where $\cI_1$ is $2\pi$-periodic in $\psi$. Thus, reduction yields
the standard form of a periodically perturbed single-degree-of freedom
system for application of Melnikov's method~\cite{M:63,GH:83}. In fact, 
inserting the series~\eqref{eq:Iseries} into the 
Hamiltonian~\eqref{eq:H0H1}, we find
\begin{align}
\cI_0 &= H_0(V,p)^{-1}(h) , \label{eq:I0}  \\
\cI_1 &= -\frac {H_1(V,p,\cI_0,\psi)}{\Omega_0(V,p,\cI_0)} . \label{eq:I1}
\end{align}

When $\mu=0$, the reduced Hamiltonian system~\eqref{eq:reduced} has a phase
portrait which coincides with that of the full system, since the vector field
is given by 
$$
\left(-\pdiff{I}{p},\pdiff{I}{V}\right) = 
\frac{1}{\Omega_0} \left(\pdiff{H}{p},-\pdiff{H}{V}\right);
$$
it therefore also has a homoclinic orbit. When $\mu > 0$, the
system~\eqref{eq:reduced} is non-autonomous, and thus we may no longer
draw a phase portrait, but we may instead construct the Poincar\'e map
on the cross section $\Sigma_{\psi_0} = \{(V,p,\psi=
\psi_0)\}$~\cite{GH:83}.  By a theorem of McGehee~\cite{M:73} the
periodic orbit at infinity, $\beta$, and its stable and unstable
manifolds $W^s(\beta)$ and $W^u(\beta)$ persist for small values of
$\mu$.

To prove transverse intersection of $W^s (\beta)$ and $W^u
(\beta)$, we apply the Melnikov method to the Poincar\'e map
$P_{\psi_0}:\Sigma_{\psi_0} \to \Sigma_{\psi_0}$ that results from
following the flow from $\psi= \psi_0$ to $\psi = \psi_0+ 2\pi$. As
noted at the beginning of Section~\ref{sec:ode_analysis}, and treated
in greater detail for the analogous Sine-Gordon problem
in~\cite{GHW:01}, a result of McGehee~\cite{M:73} allows us to apply
the Melnikov method even though the fixed point at infinity is not
hyperbolic.  The Melnikov integral can be interpreted as a normalized
distance between the stable and unstable manifolds at a specified
point on the cross section $\Sigma$.  As in~\cite{MH:88}, we may then
apply a version of the usual Melnikov method~\cite{M:63,GH:83} to the
reduced  system~\eqref{eq:reduced}. This would lead to a Poisson
bracket involving involving $\cI_0$ and $\cI_1$ in the Melnikov
integrand, but, via Equations (\ref{eq:I0}-\ref{eq:I1}), this is
equivalent to the `$p-V$' Poisson bracket of the original functions
$H_0$ and $H_1$.  This yields Theorem~\ref{thm:Mel} as stated in
Section~\ref{sec:P_inf}.

\bibliographystyle{elsart-num}
\bibliography{GWH_nls}

\end{document}